# Solvent Sensitivity of Protein Unfolding: Study of Chicken Villin Headpiece Subdomain in Water-Ethanol and Water-DMSO Mixtures

**Rikhia Ghosh, Susmita Roy and Biman Bagchi[*]**

**Solid State and Structural Chemistry Unit, Indian Institute of Science, Bangalore 560012, India**

**Email: bbagchi@sscu.iisc.ernet.in**

## Abstract

In the present work we study and compare unfolding of a small protein, chicken villin headpiece (HP-36) in two different aqueous binary mixtures, namely water-ethanol (EtOH) and water-dimethyl sulphoxide (DMSO). In both the binary mixtures, HP-36 is found to unfold (fully or partially, depending on the mixture) under ambient conditions, that otherwise requires temperature as high as ~600 K to denature in pure aqueous solvent. In all the cases, first step of unfolding is found to be similar, i.e. separation of the cluster formed by three hydrophobic (phenylalanine) residues, namely Phe-7, Phe-11 and Phe-18, which constitute the hydrophobic core, thereby initiating melting of helix-2 of the protein. Subsequent unfolding steps follow different paths in different chemical environments. As both water-DMSO and water-ethanol show composition dependent anomalies, so do the details of unfolding dynamics. With an increase of co-solvent concentration different partially unfolded intermediates are found to be formed in both the cases. This is reflected in a remarkable *non-monotonic composition dependence of several order parameters, including fraction of native contacts and protein-solvent interaction energy*. The emergence of such partially unfolded states is particularly attributed to the preferential solvation of the hydrophobic residues by the ethyl groups of ethanol and methyl groups of DMSO. While in DMSO the protein gradually attains a completely unfolded state at $x_{DMSO}$=0.30, unfolding in water-ethanol appears to be more complex and sensitive to solvent composition.

## I. Introduction

While a general picture of protein folding and unfolding is beginning to emerge using the concept of an energy-entropy funnel with intrinsic ruggedness [1, 2], the details of the kinetics and pathways are still far from our understanding at present. The reason is the absence of information of local changes involved in these transitions. Recent advances in two dimensional infrared (2D-IR) vibrational echo spectroscopic techniques, pioneered by among others, Fayer and coworkers, have developed a reliable method to measure dynamics of the proteins that is sensitive to the details of spatial arrangement of amino acid residues around a given peptide group [3-6]. 2D-IR spectroscopy allows unprecedented accuracy of time scale (in the ps range) of local motion that is required to study protein unfolding in particular, whereas conventional techniques (like NMR) are reliable only at longer time scales (in the order of ns). In this article we propose and study unfolding of a small protein with particular emphasis on melting/unfolding of secondary structure. The novelty of the present work is the use of suitable aqueous binary mixtures whose composition can be varied to induce unfolding at ambient conditions. This allows a direct comparison between experiments and theory aided by the simulations performed.

The funnel picture of protein folding was essentially introduced by Bryngelson and Wolynes in their landmark work in 1989[1]. They described a two order parameter theory to understand the dynamics from extended or unfolded state to final compact native state. Two order parameters introduced to formulate the model are essentially (i) the fraction of native contacts ($\eta$) and (ii) the radius of the biopolymer ($R$). By employing a Flory type theory for polymer collapse [7], they derived a two dimensional free energy surface of protein folding that depends on these two order parameters $\eta$ and $R$, in addition to depending on the thermodynamic parameters like temperature and solvent condition. Bryngelson and Wolyness applied a Smoluchowskii type diffusion equation to express the motion of the protein in two dimensional ($\eta$-$R$) configuration space. The free energy surface constructed by them thus contains two minima, one for the extended state characterized by small value of $\eta$ and large value of $R$, and the second one for native state distinguished by large $\eta$ and small $R$. Protein folding phenomenon was then described as transition of the protein from extended state minimum to the global native state minimum. While developing the theory, Bryngelson and Wolynes invoked the concept of "principle of minimum frustration" [1, 2]. The principle fundamentally gives the idea that, for a particular protein amino acid sequences are chosen in such a way so that native state is the most stable state. Thus the undesired amino acid interactions in the folding pathway are minimized so that formation of native state is a very smooth process, thereby minimizing the frustration that tends to create in the folding pathway. Nevertheless, even though the level of frustration is reduced naturally, some of it is left in the system due to presence of local minima in the energy landscapes of protein. The outcome of sequence dependent stabilization of native state is that free energy surface of protein folding can be viewed as 'funneled energy landscapes' that are critically directed towards the native state [8]. Thus the folding funnel landscape describes protein folding phenomenon as folding of protein to the native state through any of the possible large number of pathways and intermediates, rather than choosing a single mechanism. Although some proteins seem to show unique folding path through well-defined intermediate states, many aspects of folding picture have been accepted by the community at large. It is expected that different pathways could be accessed under different conditions, to achieve the same native state.

Recently, there has been substantial insight regarding the knowledge of native states as well as the stable intermediate states of several proteins [9-12]. However, less information is available about the path followed by protein to fold to correct, native state and the characteristics of the intermediates along the way. These intermediates are mostly short lived and thus very difficult to trap [13]. Due to formation of such unstable intermediates in the pathway of unfolding most of the studies on protein folding focus on the native state conformation rather than looking at the unfolded states [14, 15].

The energy landscape of a particular protein is perturbed by various external factors that include temperature, pressure as well as nature of solvents. Water is found to be a universal solvent for proteins, stabilizing the native state in most of the cases [16, 17]. Many computer simulation studies have been carried out with water as the solvent [18,19]. Folding pathway under such condition is dominated by hydrophobic and hydrophilic interactions among the amino acid residues of the protein and also hydrogen bonding propensity of side chains and backbone amide groups among themselves as well as with surrounding water molecules.

In contrast to the theoretical studies, experimental studies of protein folding and other kinetic studies involving enzymes seldom employ pure water, and a variety of co-solvents are being used for solvating such biomolecular systems before undergoing experiments. These co-solvents are found to play diverse roles in influencing the structural transitions of such systems. While some co-solvents are famous for stabilizing the native structures (glycerol, trimethyl amine N-oxide), some bring about denaturation in the same (urea, guanidium chloride).

Although an impressive number of studies have been devoted to explore the dynamics of water around a biomolecule [20-23] and protein unfolding/folding, relatively fewer studies have been performed on proteins in binary mixtures. Binary mixture of solvents have always served as promising ones in terms of large change in physical and chemical behaviors that they exhibit compared to the individual components. Due to striking irregularity revealed by such solutions, they have been one of the prime interests of study since decades [24, 25]. Water-ethanol (EtOH) and water-dimethyl sulphoxide (DMSO) are two such very important binary solvents, widely used in biology as essential solvents because of their unique behavior. Both water-ethanol and water-DMSO solutions are famous for exhibiting striking anomalies at various concentrations, that essentially arise due to structural transformation of co-solvent molecules through hydrophobic as well as hydrophilic interactions. Recent studies in water-DMSO have revealed that DMSO execute percolation like phase transition at concentration range $x_{DMSO} \sim 0.10$-$0.15$, through spanning cluster formation by interaction of the $CH_3$ groups as well as hydrogen bonding [26]. On the other hand, water-ethanol binary mixture is also found to exhibit abnormal behavior at various concentration ranges. Recent simulation studies have shown that at very low concentration ($x_{EtOH} \sim 0.06\text{-}0.10$) a bi-continuous phase forms by aggregation of ethanol molecules, leading to weak phase transition in the system [27]. In another significant experimental work, Jurrinen et al have performed synchrotron x-ray Compton scattering technique to show the structural transformation in solution of ethanol from a new perspective [28]. The experiments reveal two distinct structural regimes, one at low concentration ($x_{EtOH} \sim 0.05$) characterized by increase in intramolecular bond lengths, and another at high concentration range ($x_{EtOH} > 0.15$), characterized by excess density. However, the two co-solvents studied in this article (ethanol and DMSO) have vastly different characteristics regarding their protein unfolding ability at moderately high concentration. While DMSO is known to promote complete unfolding [29-31], ethanol is found to exhibit partial unfolding of proteins in general. In fact, ethanol is found to be a very well-known solvent for biomolecules. It is of course the universal co-solvent in various liquors since ages. A recent experimental work successfully provides quantitative solvation properties of lysozyme in water-ethanol binary mixtures, where preferential binding of ethanol molecules are obtained for both the native state as well as unfolded state [32]. In another work, it has been shown that fluorinated ethanol (trifluoroethanol, TFE) brings in conformational transition in proteins to form new stable conformational states that resemble 'molten globule' intermediate characterized by high α-helical content and disrupted tertiary structure [33, 34]. Various studies of biomolecules in water-ethanol have revealed that due to structural transformation from the typical tetrahedral-like water to the chain-like alcohol clusters in alcohol–water mixtures, secondary structures of proteins change from β-strand to α-helical strand near the transition concentration [35]. A recent study of the protein chymotrypsin inhibitor-2 (CI2) in aqueous ethanol shows an increase of radius of gyration up to mole fraction of ethanol $x_{EtOH} \sim 0.2$ and after that the following decreases significantly [36]. The universal feature of alcohol induced structural change of the protein from β-strand to α-helix is demonstrated in this work.

These interesting studies have driven us to explore how the structure of a small protein rearranges with adding ethanol concentration and what is the microscopic reason behind the same. In a very recent study we could identify the sequence of steps that takes place while unfolding of a protein in water-DMSO binary mixture as the DMSO concentration is increased [37]. In particular, the anomalies in the mixture around $x_{DMSO} \sim 0.15$ add interesting twist to the story.  However, no systematic theoretical study seems to exist on unfolding in water-ethanol binary mixture. As unfolding can be induced in mixed solvents by varying composition of the solute or co-solvent, it provides us with a great opportunity to study various aspects of unfolding, especially the sensitivity to varying environment and hence a glimpse of the energy landscape of the protein. For this purpose we have taken a 36 residue protein, chicken villin headpiece (HP 36) and looked at its change of dynamics with varying ethanol concentration. As mentioned, we have discussed studies of HP 36 in water-DMSO mixture in great detail elsewhere [37]; here we concentrate on water-ethanol mixture but also compare the results with those obtained for water-DMSO mixture as well as in neat water where unfolding requires high temperature. HP-36 is a small globular protein that represents the thermostable subdomain present at extreme C terminus of the 76-residue chicken villin headpiece domain [38, 39]. Villin is a unique protein which can both assemble and disassemble actin structures. HP-36 contains one of the two F-actin binding sites in villin necessary for F-actin bundling activity [40, 41]. It is a very interesting protein that although being a very small one, comprises of three α-helices as well as one compact hydrophobic core thereby bearing characteristic of large proteins. The three α-helices are denoted as helix 1 (Asp-4 to Lys-8), helix 2 (Arg-15 to Phe-18), and helix 3 (Leu-23 to Glu-32).The biological activity is believed to be centered on helix 3 [41]. Thus simulation studies with this protein are computationally less expensive. In earlier works, a model of the protein was extensively studied by constructing a hydropathy scale for the constituting amino acids, thereby exploring the correlation between energy landscape and folding topology of the same [42, 43]. Further atomistic simulation study of HP 36 in water revealed that the protein is extremely stable in water, and partially unfolded molten globule state was achieved on increasing the temperature as high as 600K [44].

The following article reveals several interesting results based on the molecular dynamics study of HP 36 in aqueous ethanol solutions of increasing ethanol concentration. We demonstrate unusual dynamical variation of structure of the protein along with change in ethanol concentration and correlate the same with the anomalous behavior of water-ethanol binary mixture at particular critical concentrations, which is brought about by the aggregation of ethanol molecules. It is found that with progressive addition of ethanol the protein goes through a structural transition, initially unfolding partially followed by refolding of the same. Time evolution of several order parameters provide us with the same anomalous structural changeover. We further provide understanding at molecular level in order to apprehend the reason for such striking features of protein at different co-solvent concentrations.

Initial steps of unfolding are found to be similar to those of unfolding in water-DMSO binary mixture as well as temperature induced unfolding, that is the separation of the hydrophobic core (instigated by the separation of Phe18 from Phe7 and Phe11). Such separation is possible due to cumulative interaction of hydrophobic core with hydrophobic part of the co-solvent molecules ($CH_3$-$CH_2$-). However, as ethanol concentration increases, distance between the phenyl alanine residues decreases considerably, that is unlike the other cases of unfolding studies. At $x_{eth} \approx 0.25$ the protein achieves a partially folded state that is native-like. We find from the corresponding snapshots that the α-helices are reformed at this stage that is similar to the results obtained for chymotrypsin inhibitor 2, which have been discussed beforehand. Thus the results obtained here confirm the propensity of ethanol molecules to bring about stabilization of α-helices. On further increasing ethanol concentration another set of abnormal structural variation of HP 36 is observed, that is again related to the localized availability of free ethanol molecules driving the partial unfolding.

The rest of the article is arranged as follows. Details of the system and simulation are discussed in section II. Section III elaborates the main results obtained in this work, that include anomalous variation of several order parameters with change of ethanol concentration, such as radius of gyration, root mean square deviation, average fraction of native contacts formed at equilibrium, comparative maps of hydrophobic native contacts at different ethanol concentration. The later part of this section provides the microscopic details of the reason behind such anomalous structural transition of the protein, which is mainly attributed to the aggregation of ethanol molecules at higher concentrations. In section IV, the results obtained are compared with that in case of water-DMSO solution. Section V presents a simple theoretical analysis of solvent dependence of unfolding that is derived combining the aspects of Bryngelson-Wolynes theory with Marcus theory of electron transfer. The brief summary of the results obtained is given in section VI.

## II. Simulation Details

Details of water-DMSO studies have been reported elsewhere [37], so we discuss only water-ethanol system details here. We have performed molecular dynamics simulation of the protein HP-36 in water ethanol binary mixture at various ethanol concentrations. All the simulations are done at 300 K temperature and 1 bar pressure. The simulations are initiated with the crystal structure of HP-36 that is obtained from NMR data of the villin headpiece Subdomain [38,39]. The coordinates of the crystal structure are collected from Protein Data Bank (PDB code 1VII). The terminal residues of the protein (Met1 and Phe36) are capped properly. We have used the extended simple point charge model (SPC/E) for water [41]. We have treated the ethanol molecules as united atoms within the GROMOS53a6 force field [46], i.e., full atomistic details have been retained except for the hydrogen atoms attached to carbon atoms. At first water-ethanol binary solution was prepared for various concentrations in cubic boxes, with sides of length 3.0 nm, although minor variations were there in all cases to maintain proper mole fraction ratio (concentrations of ethanol taken are 0.05, 0.10, 0.15, 0.20, 0.25, 0.30, 0.35 and 0.40 respectively). After performing steepest descent energy minimization, equilibration of the system was done for 2 ns keeping temperature and volume constant. After that we again performed an equilibration for 2 ns keeping pressure and temperature constant. Finally production run was executed for 20 ns in a NPT ensemble. Temperature was kept constant using Nose-Hoover Thermostat [47] and Parinello-Rahman Barostat [48] was used for pressure coupling. After preparing the binary mixtures, protein was dissolved in each of them and again same procedure was followed for energy minimization. The box size taken in this case is larger (~7 nm), to accommodate the protein as well as permitting better solvation by housing more solvent molecules in the box. The solvent was further equilibrated for 20 ns by restraining the position of protein while allowing the solvent to move. This method is popularly known as position restrained molecular dynamics simulation. Finally production run was performed for 100 ns in NPT ensemble. Periodic boundary conditions were applied and non-bonded force calculations were employed by applying grid system for neighbor searching [49]. The cut-off radius taken for neighbor list and van der Waals interaction was 1.4 nm. To calculate electrostatic interactions we used particle mesh Ewald (PME) with a grid spacing of 0.16 nm and an interpolation order of 4.

## III.  Study of Protein Unfolding in Water-Ethanol Binary Mixture

As already discussed in the Introduction section, water-ethanol binary mixture exhibits interesting anomalies over a wide range of composition which are reflected in various physicochemical properties [50, 51]. In our earlier work [27], we had shown that the anomalies in low concentration limit of water-ethanol binary mixture arise to due to emergence of unexpected microheterogenity in the system. This is attributed to the formation of ethanol clusters through hydrophobic and hydrophilic interactions between the ethyl and hydroxyl groups respectively at a concentration range $x_{eth} \approx 0.06 - 0.10$.

Now when a biomolecular system is dissolved in such mixed aqueous solution, their structural and dynamical properties are expected to change as a function of solvent composition. In fact, we found that in the concentration range as low as $x_{eth} \approx 0.05$, the collapsed state of a homopolymer gains surprising stability, while in bulk water the same chain exhibits bistability between the collapsed and extended states [27]. We further intended to see the effect of such aggregation in the mixed solvent system on a more pragmatic system such as a protein. Thus we have taken a small protein HP-36 that bears characteristics of large proteins due to the presence of 3 $\alpha$-helices and a hydrophobic core. Our objective is to explore the effect of change of ethanol concentration on the dynamic properties of the protein.

## A. Solvent Composition Dependent Structural Changes of Protein: Variation of C-α RMSD and Radius of Gyration

The standard way of analyzing structural stability of a protein in a MD simulation is to monitor the root mean square deviation (RMSD) from the initial structure along the simulation. The dynamical change of RMSD of C-$\alpha$ backbone of the protein along with change in ethanol concentration is shown in **Figure 1a**. **Figure 1b** shows the corresponding average values at each individual mole fraction. We find that on adding ethanol up to mole fraction $x_{eth} \approx 0.10$, gradual unfolding of the protein structure takes place. However on further addition of ethanol the process of unfolding is somewhat hindered as is seen from **Figure 1b** ( $x_{eth} \approx 0.10 \sim 0.20$ ). After that there is sudden drop in the value of average rmsd at $x_{eth} \approx 0.25$, signifying the transition of the structure towards folded state, which again jumps to a very high value at $x_{eth} \approx 0.30$, thus indicating the formation of most unfolded state in this case. At higher concentrations, such oscillations between the partially folded and unfolded state prevail ($x_{eth} \approx 0.35 \sim 0.40$).

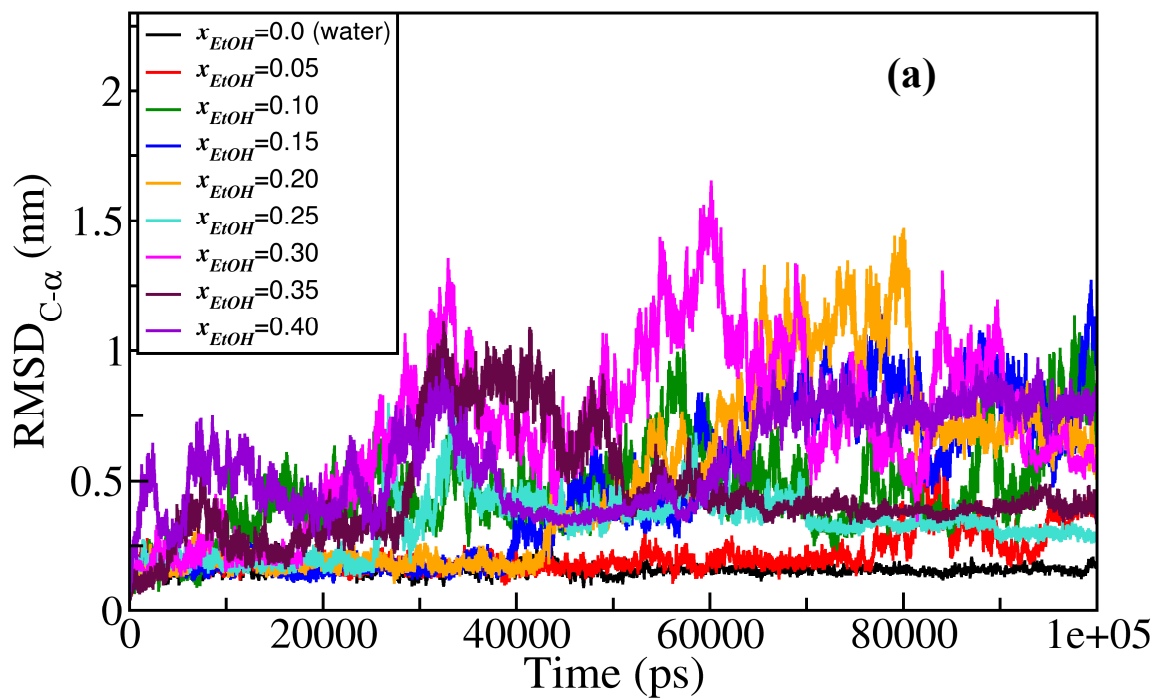

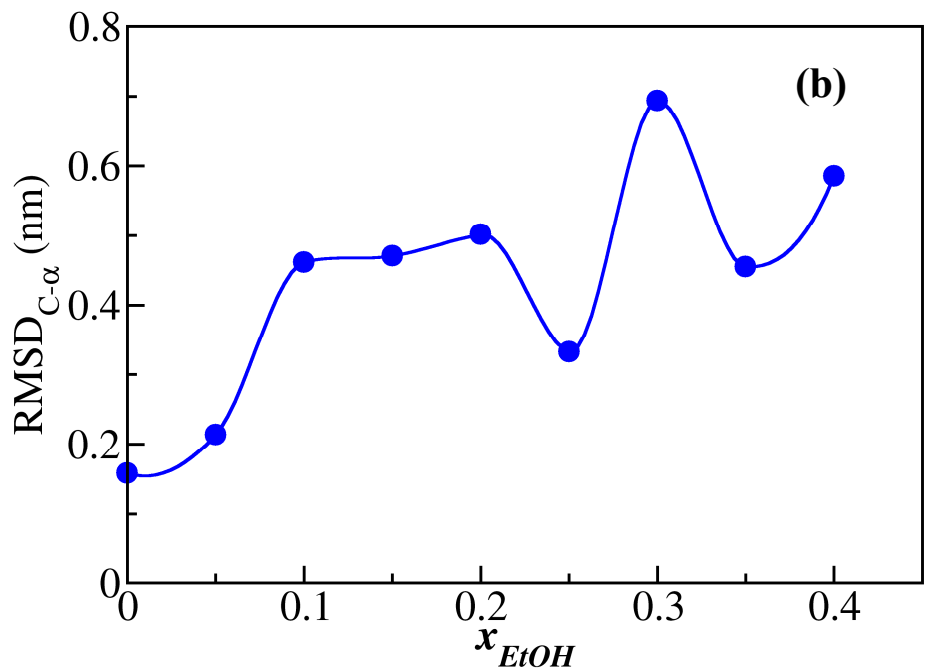

**Figure 1.** *(a) Time evolution of root mean square deviation (rmsd) of C-α backbone and (b) Average rmsd over the whole time trajectory bearing a clearer picture of solvent composition dependent structural changes.*

Although rmsd serves a good measure of the degree of conservation of a structure, it is still limited, and analysis of radius of gyration is helpful to bring further insight towards the same. We have calculated time evolution of radius of gyration of HP 36 by changing the co-solvent concentration from $x_{eth} \approx 0.0$ to $x_{eth} \approx 0.40$ which is shown in **Figure 2a**, along with equilibrium average values shown in **Figure 2b**. In this case also we find the same signature as that obtained from rmsd values, i.e., gradual unfolding up to $x_{eth} \approx 0.10$ followed by less significant change of structure in the concentration range $x_{eth} \approx 0.10 \sim 0.20$. On further increasing ethanol concentration the protein structure fluctuates between partially folded and unfolded states ($x_{eth} \approx 0.20 \sim 0.40$).

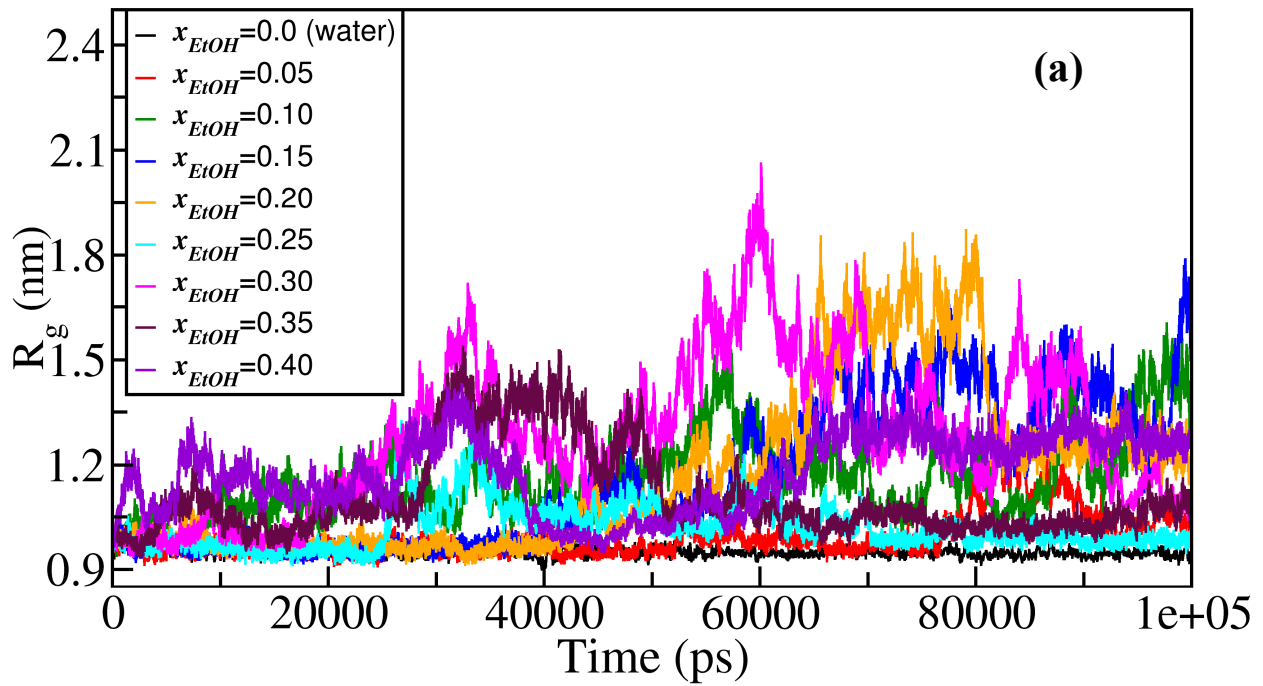

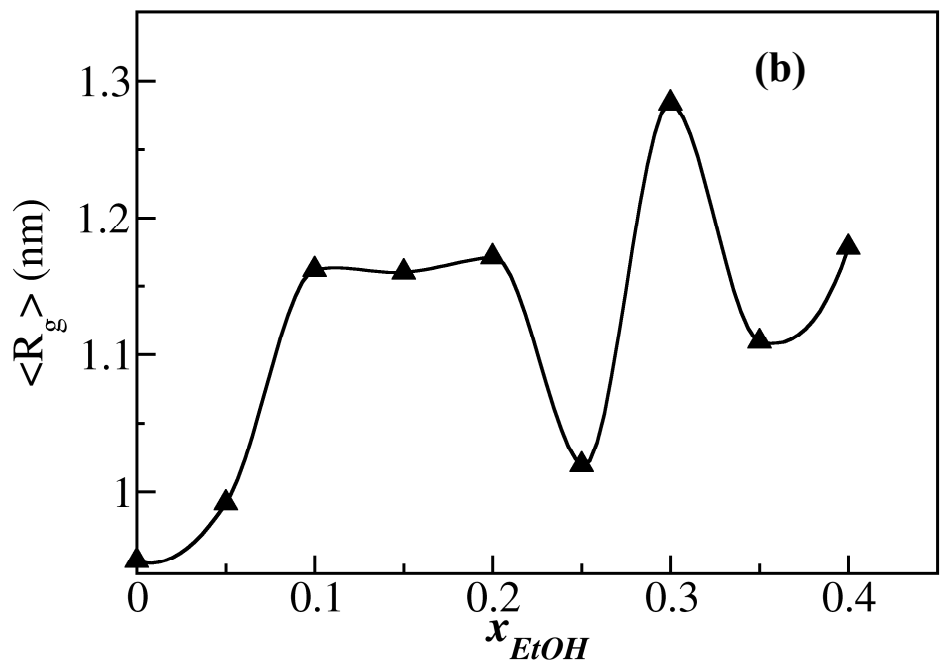

**Figure 2.** *(a) Time trajectory of radius of gyration of the protein at different mole fraction of ethanol. (b) Equilibrium average of radius of gyration over the whole trajectory plotted against increasing ethanol concentration.*

**B. Study of Fraction of Native Contacts Formed and Broken**

Formation or rupture of specific contacts between respective residues in a protein serves as an important parameter in the study of protein dynamics. Proteins are often known to fold to the correct native structure making some precise contacts among the hydrophobic and hydrophilic residues respectively. While the process of unfolding being executed, these important contacts break up in the first place making the course of unfolding smoother for the same. In order to understand the structural changes induced by increasing ethanol concentration, we have evaluated the average value of fraction of contacts ruptured or restored in the protein ($\langle \eta \rangle$) relative to that of the native, correctly folded structure (**Figure 3**). The cut off for formation of native contacts between the residues has been taken as 4 $\overset{\circ}{\mathrm{A}}$ as per the standards.

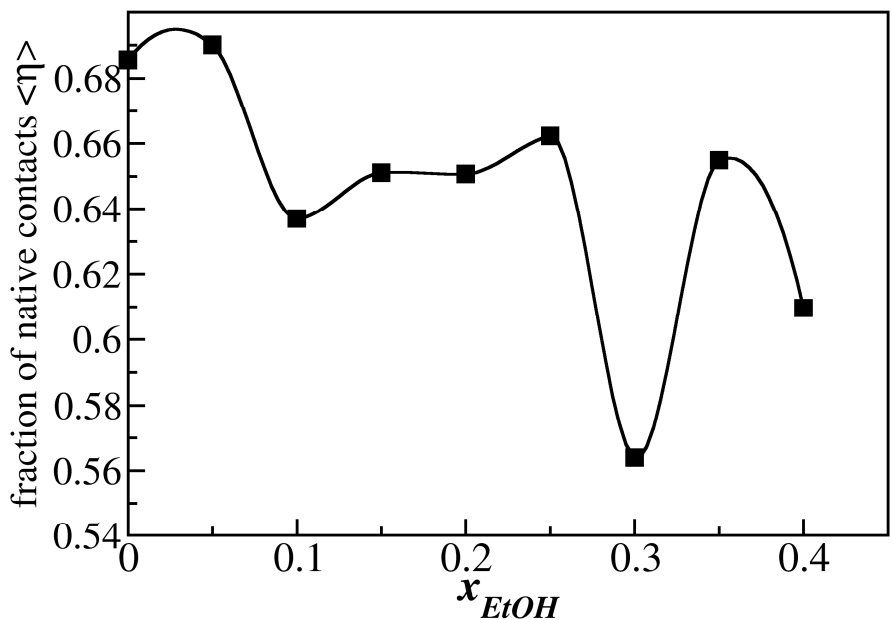

**Figure 3.** *Equilibrium average of fraction of native contacts as a function of increasing ethanol concentration.*

In this case we find that on adding a very small quantity of ethanol ($x_{eth} \approx 0.05$) the relative number of native contacts formed in the protein changes a little bit compared to the number of contacts formed in bulk water. However, on increasing the ethanol concentration to $x_{eth} \approx 0.10$, the average number of fractional contacts decrease significantly, indicating the emergence of partial unfolding of the structure. The important contacts are again restored on further increase of ethanol concentration ($x_{eth} \approx 0.10 \sim 0.25$), which signifies the phenomenon of refolding. At $x_{eth} \approx 0.30$, there is huge decrease in the value of $\langle \eta \rangle$, followed by oscillation between a higher and lower value of the same at higher ethanol concentration ($x_{eth} \approx 0.35 \sim 0.40$), that is eventually a measure relative folding and unfolding of the structure.

**C. Snapshots of the HP-36 at Different Ethanol Concentration**

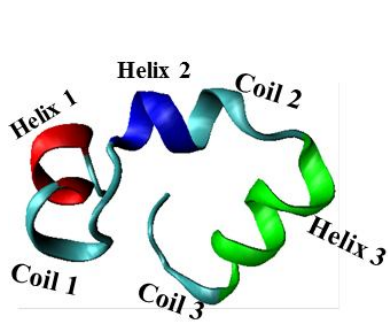 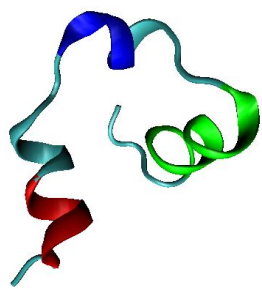 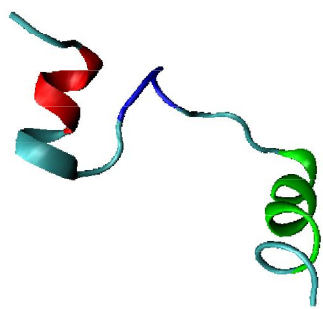

(1) Native State       (2) $x_{eth} \approx 0.05$       (3) $x_{eth} \approx 0.10$

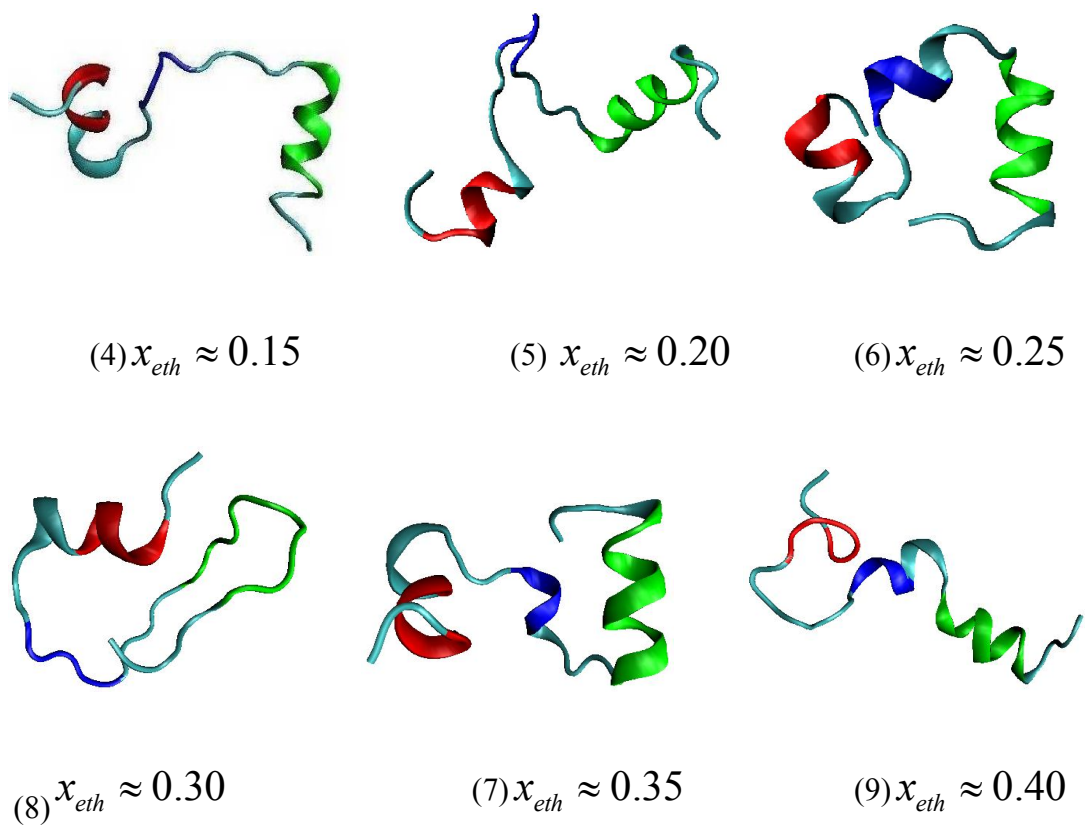

(4) $x_{eth} \approx 0.15$       (5) $x_{eth} \approx 0.20$       (6) $x_{eth} \approx 0.25$

(8) $x_{eth} \approx 0.30$       (7) $x_{eth} \approx 0.35$       (9) $x_{eth} \approx 0.40$

**Figure 4.** *Equilibrium snapshots of HP-36 at different mole fractions of ethanol. Three different helices are assigned three different colors, namely, Helix-1 (4-8 amino acid residues) is assigned red, Helix-2 (15-18 amino acid residues) is assigned blue and Helix-3 (23-32 amino acid residues) is assigned green. Coil-1, Coil-2, Coil-3 and Coil-4 are marked by cyan color. Snapshot (1) demonstrates native structure of the protein in water, (2) at $x_{eth} \approx 0.05$ (3) at $x_{eth} \approx 0.10$ (4) at $x_{eth} \approx 0.15$ (5) at $x_{eth} \approx 0.20$ (6) at $x_{eth} \approx 0.25$ (7) at $x_{eth} \approx 0.30$ (8) at $x_{eth} \approx 0.35$ and (9) at $x_{eth} \approx 0.40$.*

In order to see the structural changes induced by varying concentration of ethanol, we follow the snapshots of the protein during different stages of simulation. In **Figure 4** we provide the structures obtained after equilibrium is reached in the system. We find that at $x_{eth} \approx 0.05$, partial unfolding is initiated by disruption of Helix 2 region, as well as deformation of hydrophobic core that comprises of Phe-7, Phe-11 and Phe18. Same initial pathway towards unfolding was found to be followed by the protein in case of earlier simulation studies [37]. In the ethanol concentration range of $x_{eth} \approx 0.10 - 0.20$, much structural variation is not observed. In all the cases we find melting of second helix accompanied by melting of coil-2 and coil-3, along with slight deformation of first helix at around $x_{eth} \approx 0.20$. This is also found from the radius of gyration and RMSD plots (**Figure 1** and **Figure 2**). Inconsistency arises at $x_{eth} \approx 0.25$, where the protein is again found to acquire a native like structure, with reformation of Helix-2. This remarkable result is highly unexpected as with increasing ethanol concentration protein is expected to unfold gradually as a result of increasing protein-solvent hydrophobic interaction. Further increase of ethanol concentration leads to formation of different partially unfolded intermediates. At $x_{eth} \approx 0.30$, Helix-2 and Helix-3 melt, whereas Helix-1 remains intact. At $x_{eth} \approx 0.35$, all the three helices undergoes partial melting, and at $x_{eth} \approx 0.40$, Helix-1 melts along with partial melting of Helix-2, keeping Helix-3 intact. These unpredictable structural fluctuations provide highly interesting results, and perhaps this work reports first ever systematic study to show such fluctuating structural changes along with increasing co-solvent concentration to the best of our knowledge. These initial results demands further detailed analysis of the system that is discussed in the next sections.

**D. Study of Structural Change of HP-36 through Contact Map**

From the previous discussions on dynamical variation of HP 36, it is realized that the pathway of unfolding can be envisaged by a number of partially unfolded intermediates. This can be interpreted as the presence of a number of local free energy minima along the pathway of unfolding. To better understand the nature of these intermediates, we map the hydrophobic contacts present in the protein at each ethanol concentration after equilibrium is reached and compare them with the corresponding native hydrophobic contacts (**Figure 5**)

**(1)**

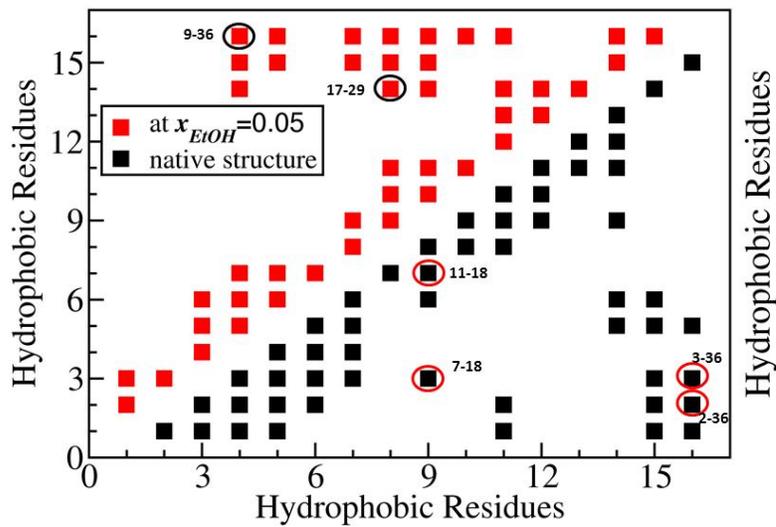

**(2)**

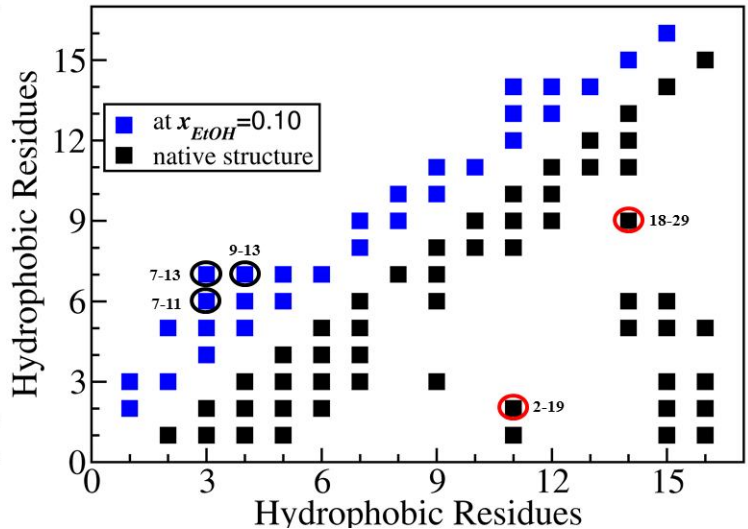

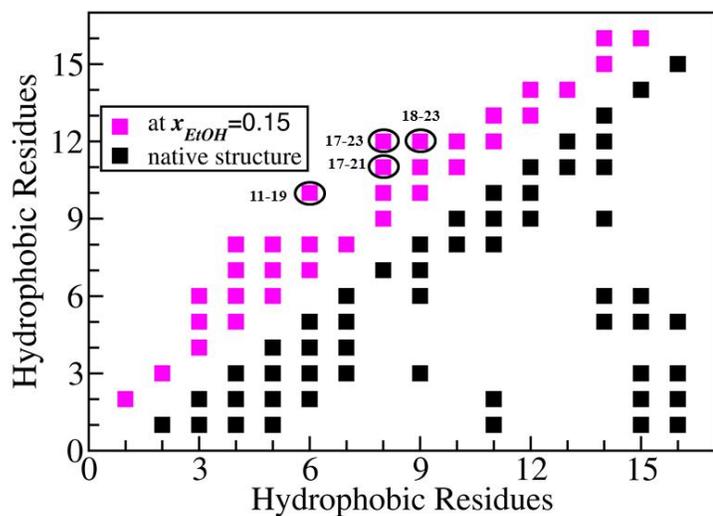

**(3)**

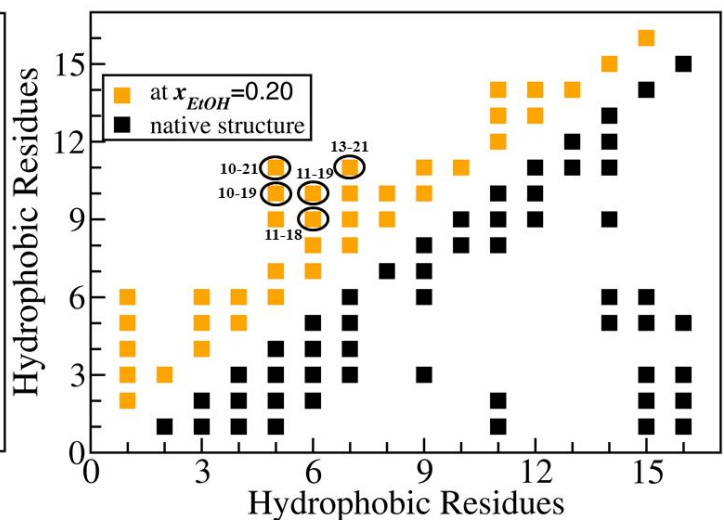

**(4)**

**(5)**    **(6)**

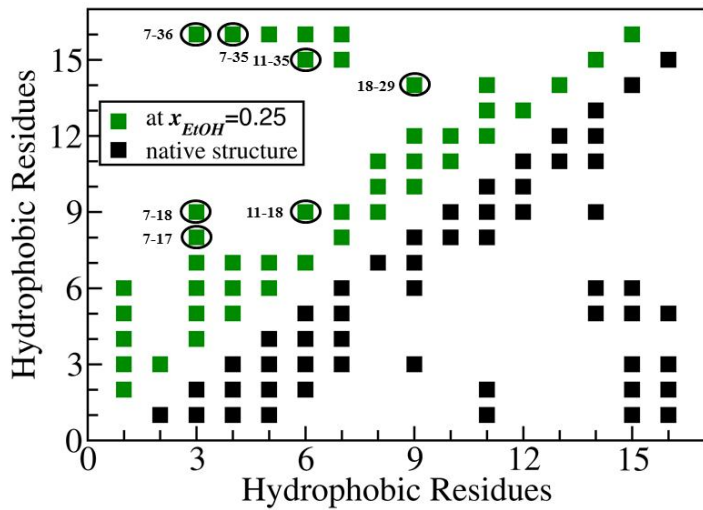 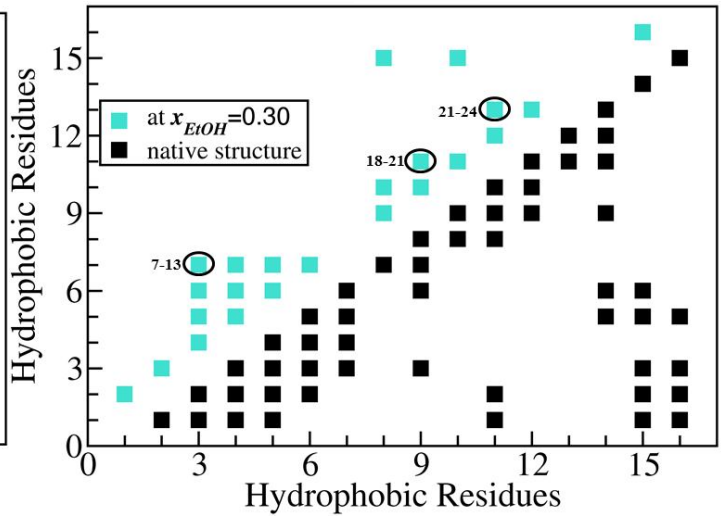

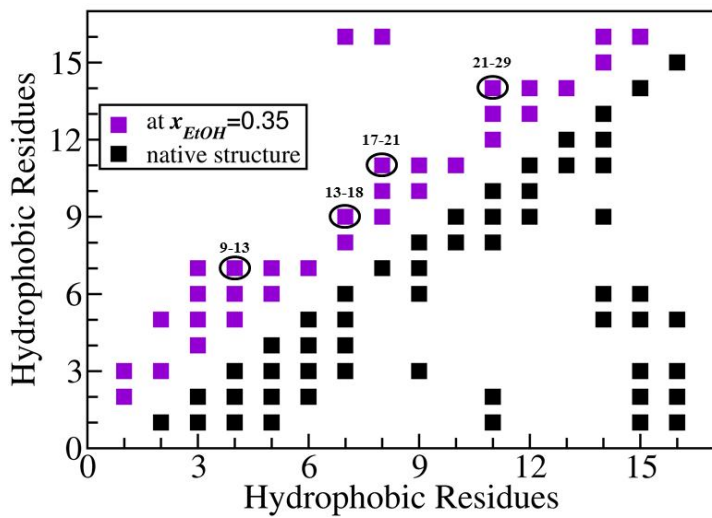 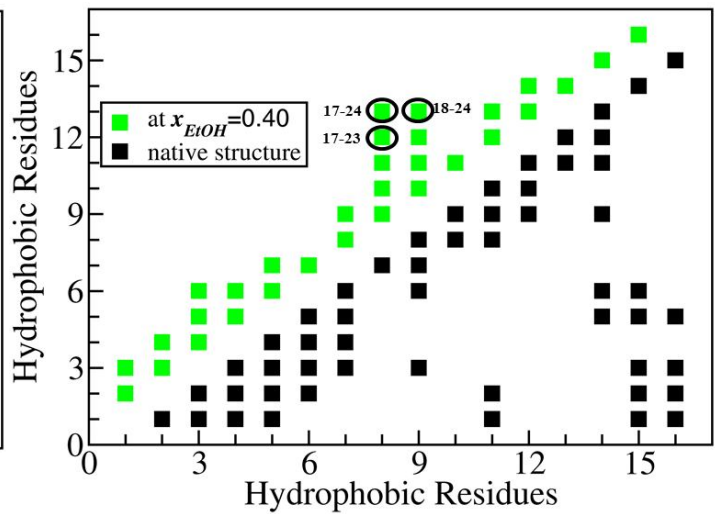

**(7)**    **(8)**

**Axis Index**

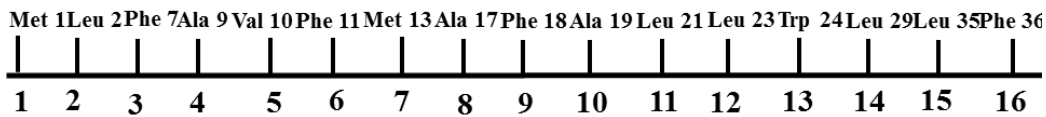

**Hydrophobic Residue**

**Figure 5.** *Hydrophobic contact mapping of HP-36 at different mole fractions of ethanol showing evolution of new contacts and breakage of old contacts, along with comparison with the native hydrophobic contacts present at equilibrium (right diagonal). The axis index given above shows the hydrophobic residue number. It is to be noted that deformation of second helix starts with separation of hydrophobic core (Phe-7, Phe-11 and Phe-18). As the protein refolds again ( $x_{eth} \approx 0.25$ ), these contacts are restored once more.*

A close inspection of **Figure 5** reveals that also at very low concentration of ethanol ( $x_{eth} \approx 0.05$ ) the tertiary structure starts breaking with the removal of long-distant contacts, marked by the removal of points placed at the farthest corner (1-36, 2-36, 1-6). The initial steps of unfolding involve separation of the two significant hydrophobic contacts (7-18 and 11-18). An important point to be noted is that in case of study of HP-36 in water-DMSO also the initial steps of unfolding were found to be the separation of hydrophobic core through the separation of Phe7-Phe18 and Phe11-Phe18 contacts [33]. In fact in case of $x_{eth} \approx 0.05$ some new tertiary contacts are found to be formed that is not present in the native structure, which implies the formation of some stable partially folded intermediate with different type of contacts present in the system (e.g., 9-36, 17-29).

At $x_{eth} \approx 0.10$, all the tertiary contacts are lost along with loss of some secondary contacts (e.g., 2-19, 18-29). This signifies further unfolding of the protein. As ethanol concentration is increased from $x_{eth} \approx 0.15$ to $x_{eth} \approx 0.20$ it is found that Phe11-Phe18 contact reappears, with the recurrence of some secondary contacts. This is in accordance with the distribution of distances between these two residues shown in the next section. From **Figure 1(b)** we find that in this region the average $R_g$ value of the protein remains almost constant. This result is also consistent with that obtained for HP-36 in water-DMSO solution at the same concentrations. However, at $x_{eth} \approx 0.25$ it is seen that almost all the secondary and tertiary contacts are restored, thereby establishing the fact that at this particular concentration the protein is again folded to a give a native-like structure. Here we find that the all the hydrophobic core contacts are reformed (7-18 and 11-18), important tertiary contacts also reappear (7-36, 11-35). At $x_{eth} \approx 0.30$ protein acquires a significantly extended state, designated by least number of contacts present in the corresponding map. At $x_{eth} \approx 0.35$ some secondary and tertiary contacts re-develop, that is also evident from **Figure 4** (second helix is reformed to some extent). At $x_{eth} \approx 0.40$, no tertiary contacts are present and formation of a partially unfolded structure takes place.

## E. Breakdown of Tertiary Structure: Initial Separation Followed by Aggregation of Hydrophobic Core

HP-36, although being a small protein (36 amino acid residues) consists of a central hydrophobic core that comprises of three phenyl alanine groups (phe-7, phe-11 and phe-18). As seen from the contact map of the native structure, these three hydrophobic residues form important tertiary contacts that contribute in correct folding of the protein. Since protein unfolding is universally observed to be associated with structural changes in hydrophobic cores, we follow the distance distribution of the three hydrophobic pairs, namely, phe7-phe18, phe11-phe18 and phe7-phe11, along the course of changing ethanol concentration (**Figure 6**). We find that all the pairs are in close contact in water. However, as ethanol concentration increases these tertiary contacts start disappearing ($x_{eth} \approx 0.10$) for the two hydrophobic pairs phe7-phe18 and phe11-phe18. In distance distribution plots we find that the first peak almost vanishes, and a second peak at a larger distance starts emerging. This phenomenon essentially signifies the breakage of tertiary structure that is also reflected in the snapshot of the protein at $x_{eth} \approx 0.10$, where the disruption of the second helix is found to be initiated. On further increasing ethanol concentration ($x_{eth} \approx 0.15 - 0.20$), to our utter surprise, these hydrophobic tertiary contacts start reappearing, that is manifested in gradual disappearance of the second peak at larger distance and recurrence of first peak at smaller distance. Thus we state that initial melting of the protein is associated with initial separation followed by nucleation of the hydrophobic core. In fact, we have found the same phenomenon happening in case of water-DMSO also, in this particular concentration range, which we will discuss in detail later. The formation of phe7-phe18 and phe11-phe18 contacts continue up to $x_{eth} \approx 0.25$, where the distribution again produces a single peak at a small separation value. Accordingly we find from the snapshots that the protein refolds at this concentration, giving average $R_g$ and RMSD values very close to that in water. This produces an astonishing result, as mixed solvents are known to substantiate unfolding process at higher co-solvent concentration. On further increasing ethanol concentration ($x_{eth} \approx 0.30 - 40$) we again observe appearance of bimodal distribution with oscillatory values of the respective maxima, indicating corresponding appearance of partially unfolded states. On the other hand, the distance distribution of phe7-phe11 does not produce any such anomalies at lower concentrations, with the first peak height decreasing gradually along with the formation of a second peak continuing up to $x_{eth} \approx 0.20$. At $x_{eth} \approx 0.25$ the height of first peak increases, followed by a small decrease of the same at $x_{eth} \approx 0.30$. After that the peak height again decreases, giving a bimodal distribution at $x_{eth} \approx 0.40$. To further verify the connection between unfolding pathway and that of the hydrophobic core, we plot the average distance of the three hydrophobic pairs respectively. In this case also we find the same trend, thereby further strengthening the fact.

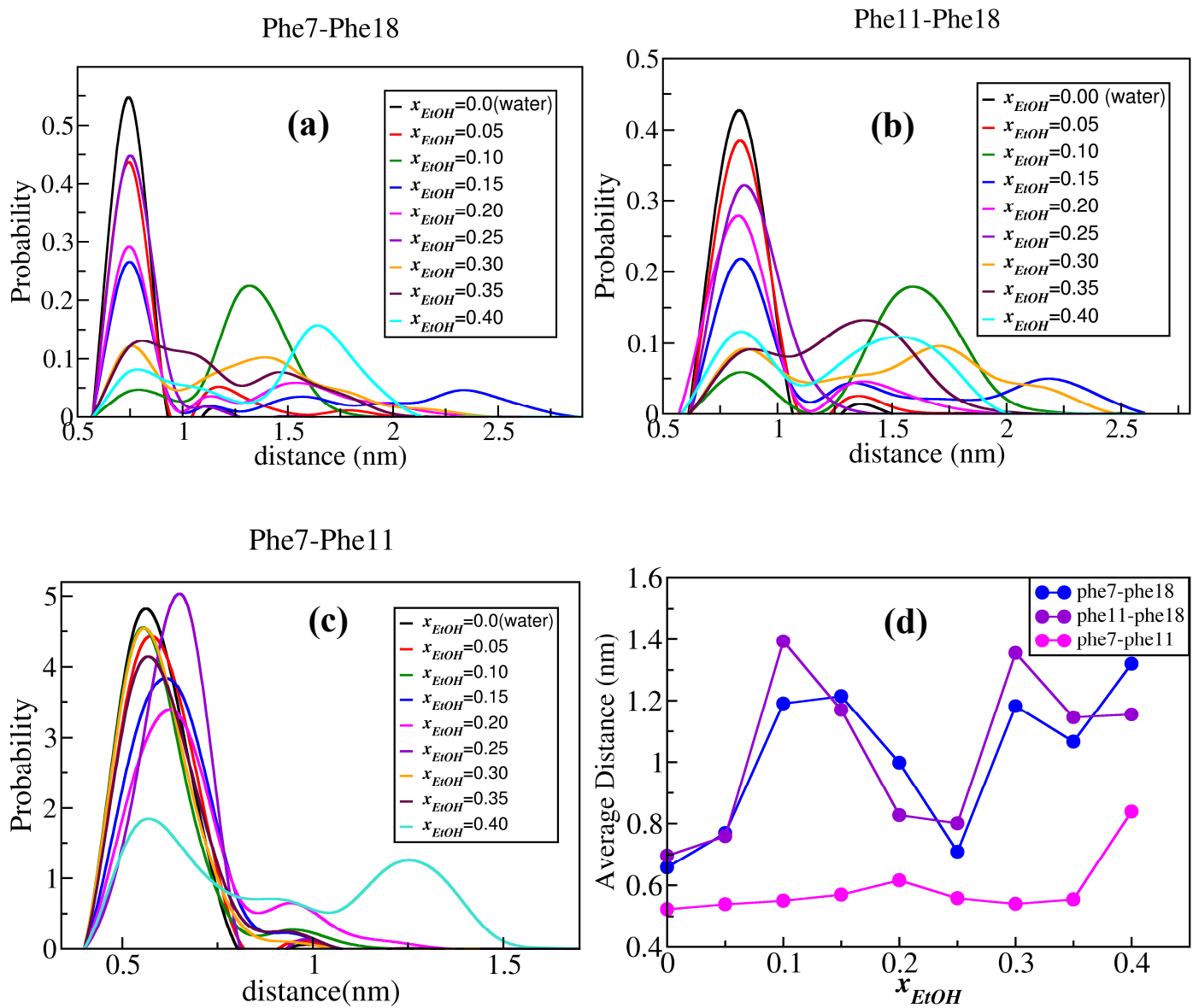

**Figure 6.** *Comparison of probability distribution of distances between the residues of hydrophobic core (comprising of phenyl alanine 7, phenyl alanine 11, phenyl alanine 18) with changing ethanol concentration. Distance distribution of (a) Phe7-Phe18 pair (b) Phe11-Phe18 pair (c) Phe7-Phe11 pair. (d) Average value of distances between these three consecutive pairs along with change of ethanol concentration.*

**F. Interaction Energies: Understanding the Role of Solvent in Protein Unfolding**

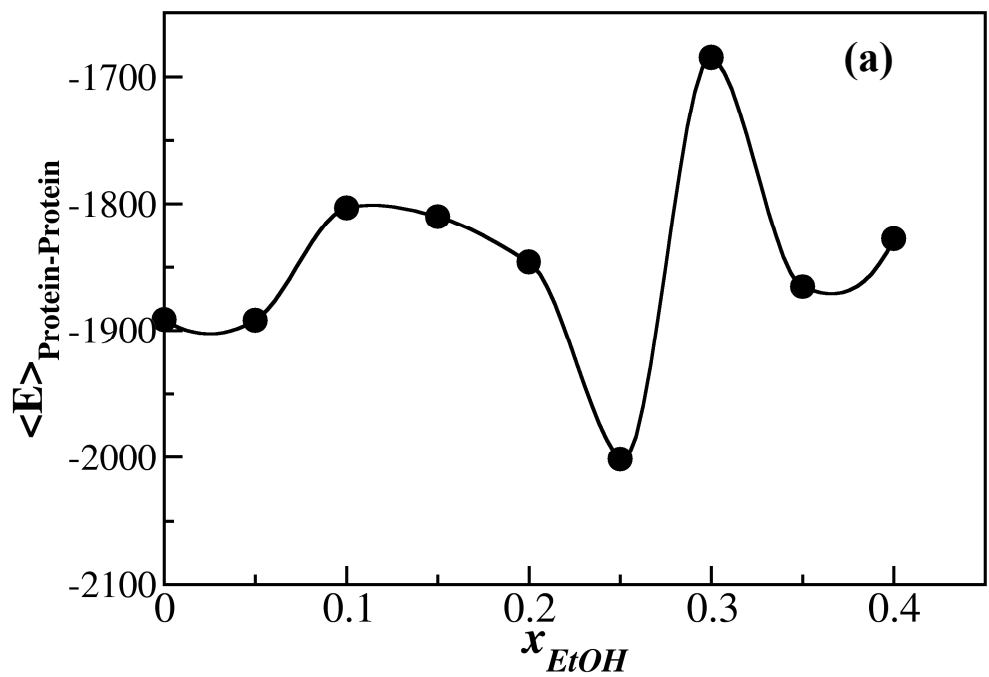

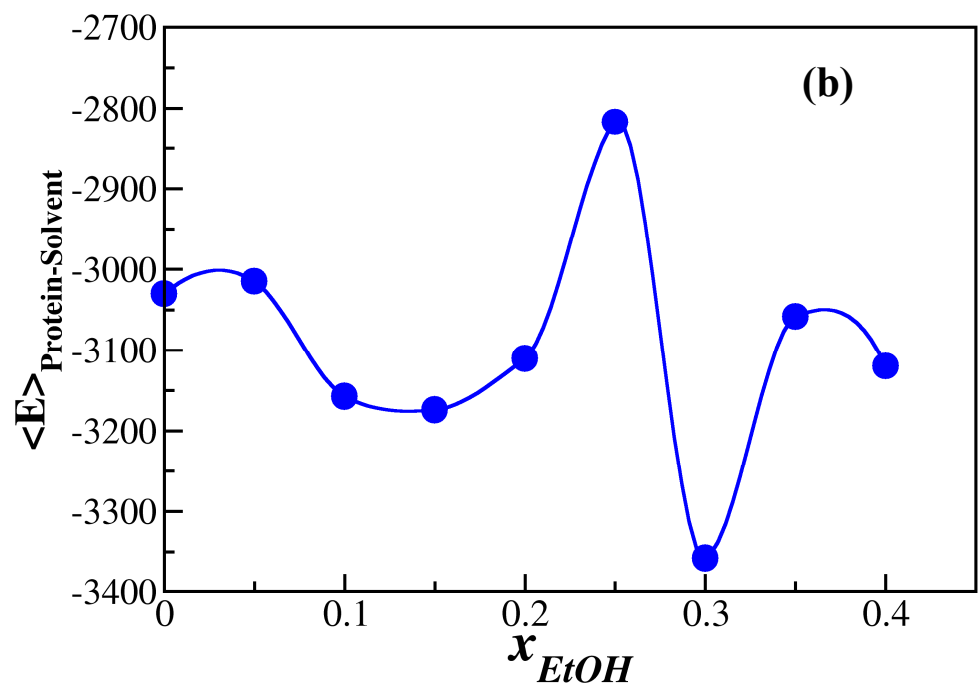

**Figure 7.** *Change of average (a) protein-protein and (b) protein-solvent interaction energy with change in ethanol concentration. Non-monotonous behavior of both protein-protein and protein-solvent interaction energy indicate dynamic structural fluctuation of the protein between partially folded and unfolded states.*

To understand the sensitivity of protein unfolding transition towards the change of solvent concentration, it is important to see how the interaction energy between the two varies along the path. We follow average of protein-protein interaction energy (**Figure 7a**) as well as average protein-solvent interaction energy (**Figure 7b**) over the whole time trajectory. It can be observed that on adding small amount of ethanol ($x_{eth} \approx 0.05$) in water, both protein-protein and protein-solvent interaction energy show insignificant change. However on increasing the ethanol concentration to $x_{eth} \approx 0.10$ protein-protein interaction energy decreases thereby destabilizing the protein. On the other hand, protein-solvent interaction energy increases indicating the initialization of unfolding of the protein. In the range of ethanol concentration $x_{eth} \approx 0.15 - 0.20$, protein-protein interaction energy increases to a small extent and protein-solvent interaction energy decreases accordingly. This means that the unfolding that started by adding a small amount of co-solvent now ceases to occur. At $x_{eth} \approx 0.25$ protein-protein interaction energy increases enormously, leading to huge decrease of protein-solvent interaction energy signifying the formation of a folded structure that is also observed in other results. On further increase of ethanol concentration, at $x_{eth} \approx 0.30$ the structural transition remarkably changes by stabilizing a highly unfolded state, signaling the very high average value of protein-solvent interaction energy and corresponding smaller value of protein-protein interaction energy. At $x_{eth} \approx 0.35 - 0.40$, both the average energies again achieve a moderate value, that is close to the energy values obtained at $x_{eth} \approx 0.15 - 0.20$, thereby marking the appearance of a partially unfolded state all over again.

All the results obtained in the preceding sections indicate a very interesting phenomena happening with increase of ethanol concentration. Obviously such unique observation demands explanation at a molecular level. In the next section, we try to develop the plausible rationalization of the same through deeper understanding of the problem.

## G.1. Study of Water-Ethanol Binary Mixture at Different Ethanol Concentration: Ethanol molecules Associate through Hydrogen Bonding at Moderate Concentration of Ethanol in Solution

To perceive a clearer picture of the phenomena at a molecular level, we looked at the equilibrated snapshots of the box. A very interesting scenario came out from this study. We have found that at lower concentration of ethanol ($x_{eth} \approx 0.05 - 0.20$), 2EtOH.1H$_2$0 clusters are prevalent in the system, formed by hydrogen bonding between hydrogen of water and oxygen of ethanol (**Figure 8 (a)**). However at around $x_{eth} \approx 0.15 - 0.20$ ethanol molecules are seen to gradually come closer to each other and aggregate around the protein. This phenomenon is essentially a local one, bringing in microheterogeneous phase separation in the system. Through a closer inspection of the system in this particular concentration range, we find the emergence of more number of aggregated ethanol molecules, comprising of three to five units of ethanol (**Figure 8(b)** and **(c)**). At $x_{eth} \approx 0.25$ it is found that the ethanol molecules are clustered together forming three to five coordinated species, which are connected to each other through ethanol-ethanol hydrogen bond. We also find that 1EtOH.2H$_2$0 species get numbered, with a very few present in the bulk and near the surface of the aggregated ethanol molecules. The structural transformation taking place in water ethanol solution at moderately high concentrations, is distinctly different from that occurring at low concentration of ethanol [24], and is already been reported in several works, as discussed in the introduction section. At $x_{eth} \approx 0.35 - 0.40$ similar molecular arrangements of ethanol are found to persist. We have also looked into the snapshots of the neat water-ethanol binary mixture at different concentration, to make sure whether such self-association of ethanol is induced by protein. There also we find same aggregation phenomena taking place with increase of ethanol concentration.

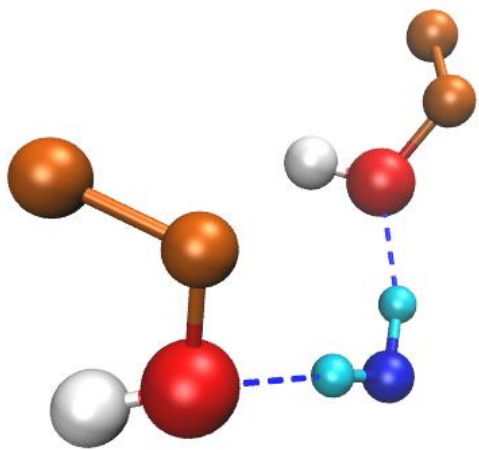

**(a)**

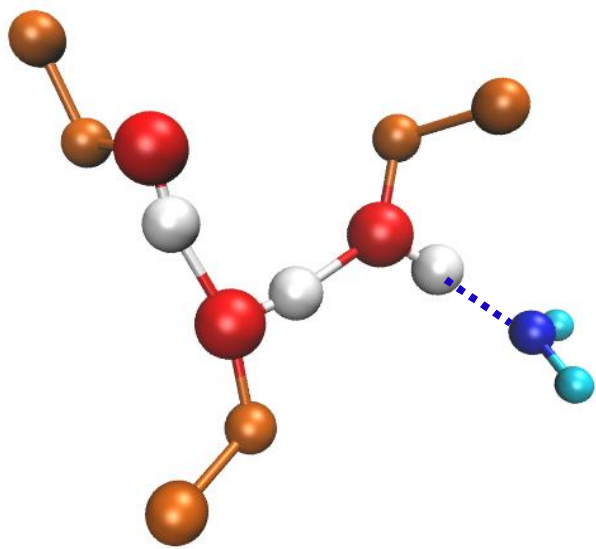

**(b)**

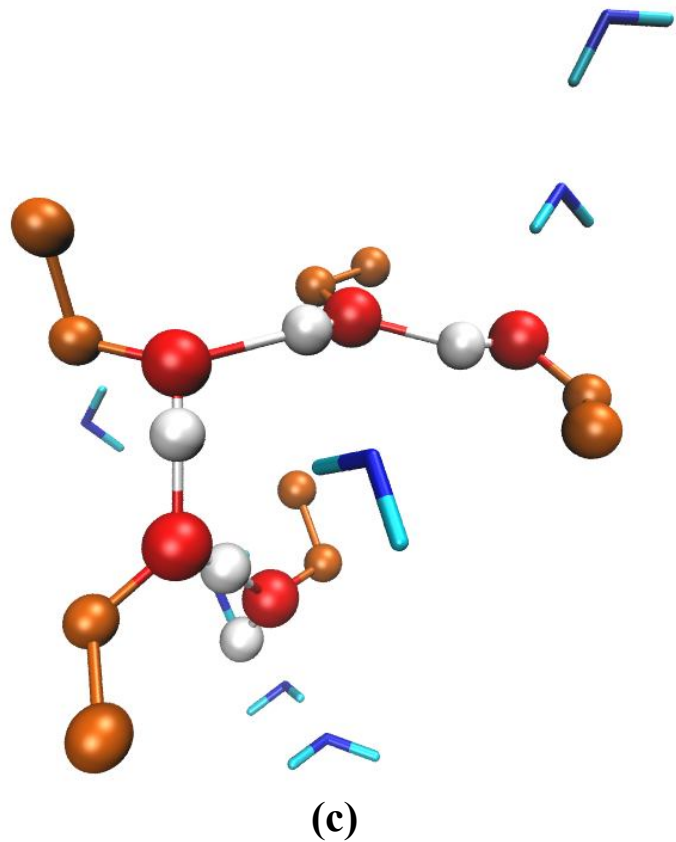

**(c)**

**Figure 8.** *Structural formations in water ethanol binary mixture observed at different concentrations of ethanol. The bronze colored balls demonstrate ethyl groups; red ones show oxygen and white balls are hydrogen atoms of ethanol molecules. The dark blue balls and lines are oxygen and light blue ones are hydrogen of water respectively.(a) $1EtOH.2H_2O$ unit formed by hydrogen bond between hydrogen of water and oxygen of ethanol. These species are abundant in the solution at low concentration of ethanol (b) Association of three ethanol molecules through hydrogen bonding between oxygen of ethanol molecule and hydrogen of another ethanol molecule. These species start forming once the ethanol concentration is gradually increased (c) Five-coordinated ethanol molecules that are ubiquitous at and onwards $x_{eth} \approx 0.25$. In this concentration range, species (a) is scarce and solution is rich of (b) and (c). The water molecules which are hydrogen bonded to ethanol are represented by ball and stick, while the free ones are indicated by dynamic bonds.*

It is rather difficult to quantify the presence of such aggregated molecular species in solution due to the microscopic scale of the problem. However, we have calculated the average number of water-ethanol and ethanol-ethanol hydrogen bonds in neat water-ethanol binary mixture (**Figure 9 (a)**). We indeed find that with increase in ethanol concentration average number of water-ethanol hydrogen bonds decrease significantly accompanied by marked increase in number of ethanol-ethanol hydrogen bond. Thus our observations are consolidated. We have also plotted the diffusion co-efficient of ethanol as a function of ethanol concentration to see whether such aggregation of ethanol molecules has any effect on the system as a whole (**Figure 9(b)**). We find that at concentration range $x_{eth} \approx 0.25 - 0.30$ diffusion co-efficient decreases by a significant amount followed by further increase at $x_{eth} \approx 0.35 - 0.40$. Thus there exists extensive anomaly in the concentration range of $x_{eth} \approx 0.25 - 0.30$, that can be attributed to the formation of three coordinated and five coordinated ethanol clusters, thus making the diffusion of the ethanol molecules through the system a very slow process. The reason for anomalous values of diffusion co-efficient at low concentration is already explained in reference [23].

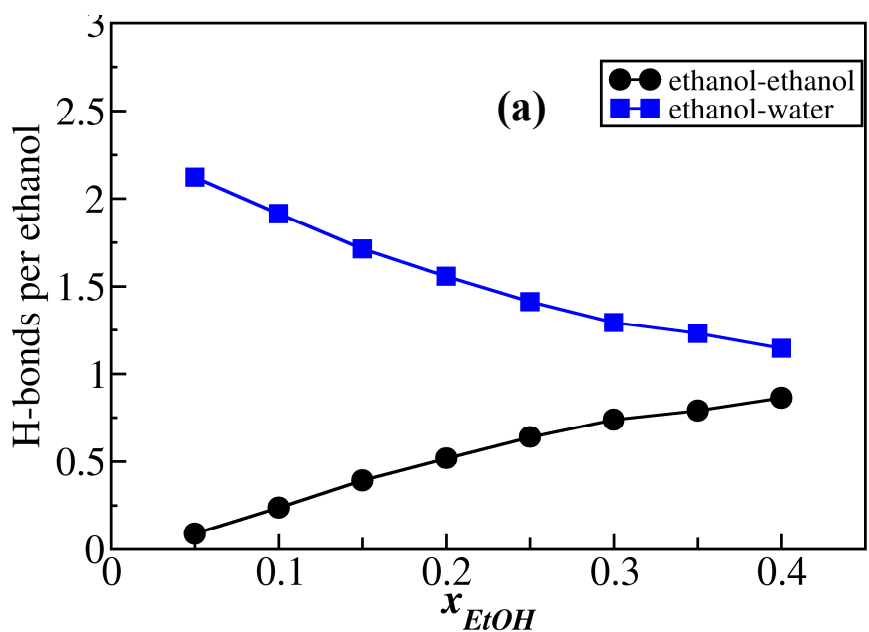

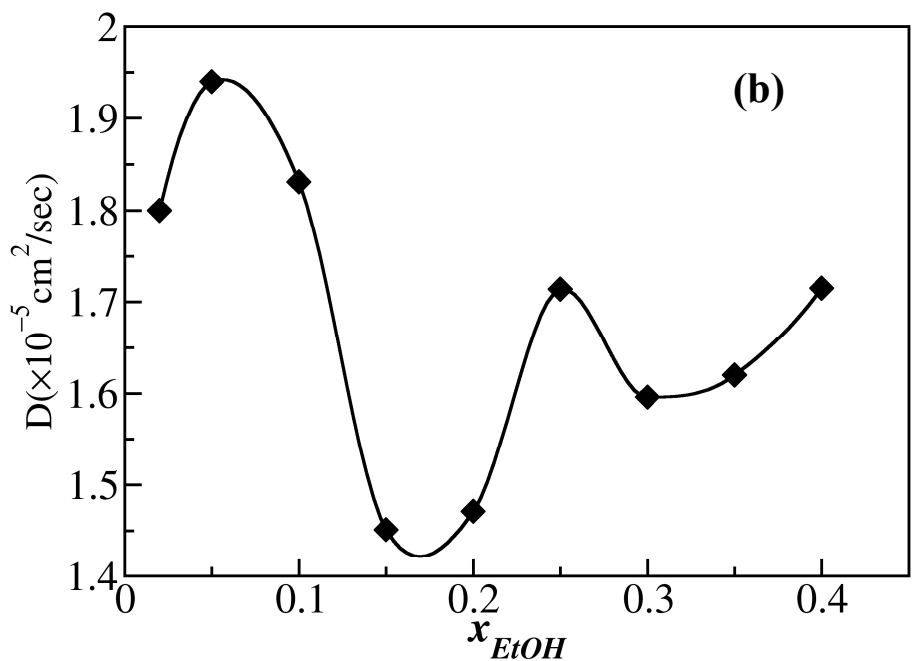

**Figure 9.** *(a) Average number of water-ethanol and ethanol-ethanol hydrogen bonds per ethanol molecule as a function of increasing ethanol concentration. Association of ethanol molecules through hydrogen bond takes place at the expense of water-ethanol hydrogen bonded species with increasing ethanol concentration (b) Anomalous variation of diffusion co-efficient of ethanol with varying concentration. The decrease of diffusion co-efficient value at $x_{eth} \approx 0.30$ strongly supports the phenomena of aggregation of the co-solvent molecules.*

## G.2. Aggregation of Ethanol Molecules Serves as the Main Reason behind the Anomalous Structural Variation of Protein

Clustering of ethanol molecules with increasing concentration alone explains most of the riddles that we faced in the abnormal structural variation of the protein. When free ethanol molecules are present in the solution, they serve as a good medium of interaction with the protein through strongly hydrophobic ethyl groups, resulting in separation of the hydrophobic core as well as initialization of meting of second helix. Now with increasing ethanol concentration free ethanol molecules become less available due to self-association. Thus the structural change in protein does not vary much in the concentration range of $x_{eth} \approx 0.15 - 0.20$. However, the refolding of the protein at around $x_{eth} \approx 0.25$ is attributed to the fact that, formation of ethanol-ethanol cluster comes to saturation in this level, thereby making very less number of free ethanol molecules available. Thus the environment around protein becomes water-like as a whole, resulting in a partially folded state of the same. On adding a little more ethanol to the solution ($x_{eth} \approx 0.30$), suddenly some free ethanol molecules are accommodated in the system, which again promotes hydrophobic solvation of the protein, thereby destabilizing the system and generating a highly unfolded state. At higher concentration of $x_{eth} \approx 0.35 - 0.40$, again the ethanol clustering starts forming in the system, making free ethanol molecules less abundant, thus stabilizing partially unfolded structures. In this context, it should be mentioned that the partially unfolded structures obtained at each concentration are quite different from each other, with the partially melted third helix and stable first helix at some concentration and vice versa in others. Presence of a number of such stable partially unfolded states can be attributed to the fact that, in protein-solvent system a large number of local minima are encountered by the protein while traversing the pathway towards unfolding. In binary mixtures, due to diverse interactions of the protein and co-solvents these minima are not separated by large free energy barriers. In such cases small change in composition of the solvent is prone to bring about large conformational fluctuation in the protein, which gets easily trapped in such local minima. Such incident is expected to occur in this particular system, thereby stabilizing different conformational forms of protein at different concentrations. However, further detailed analysis on the system is being carried out to achieve a more quantitative understanding.

In the next section we compare and contrast the results for HP-36 in water-ethanol with that in water-DMSO, and discuss how the structural changes for the two systems can be correlated.

## IV.   Unfolding of HP-36 in Water-DMSO: Comparison with the Results of Water-Ethanol

We have discussed several aspects of unfolding dynamics of HP-36 in water-DMSO binary mixture in a recent contribution [37]. DMSO concentration dependent dynamical evolution of protein trajectories from the native state to the unfolded state was thoroughly studied, along with time evolution of native contact pair formation and sequence dependent contact distance. The pathway of unfolding has been found to follow a smooth dependence on DMSO concentration, with $x_{DMSO} = 0.30$ incorporating complete unfolding of the protein (**Figure 10**).

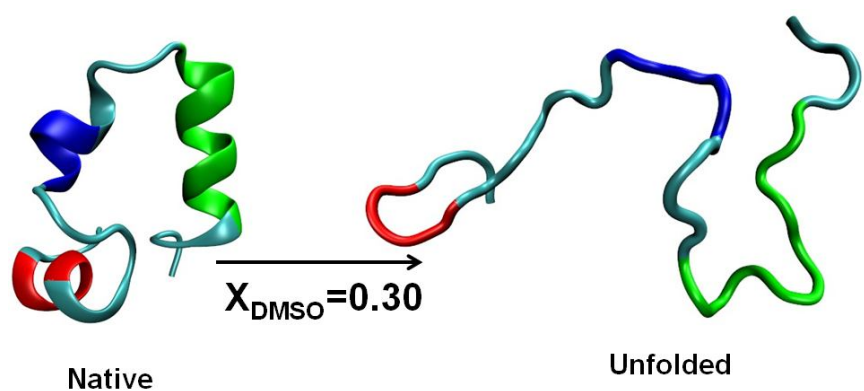

**Figure 10.** *Complete unfolding of HP-36 in water-DMSO at* $x_{DMSO} = 0.30$ *starting from the native state.*

A number of metastable states have been detected while investigating the intermediates formed during the course of unfolding, and they are found to be similar with the intermediates formed during thermal denaturation process of the same protein [44]. A molecular mechanism of DMSO induced unfolding process based on protein-solvent interaction has also been demonstrated.

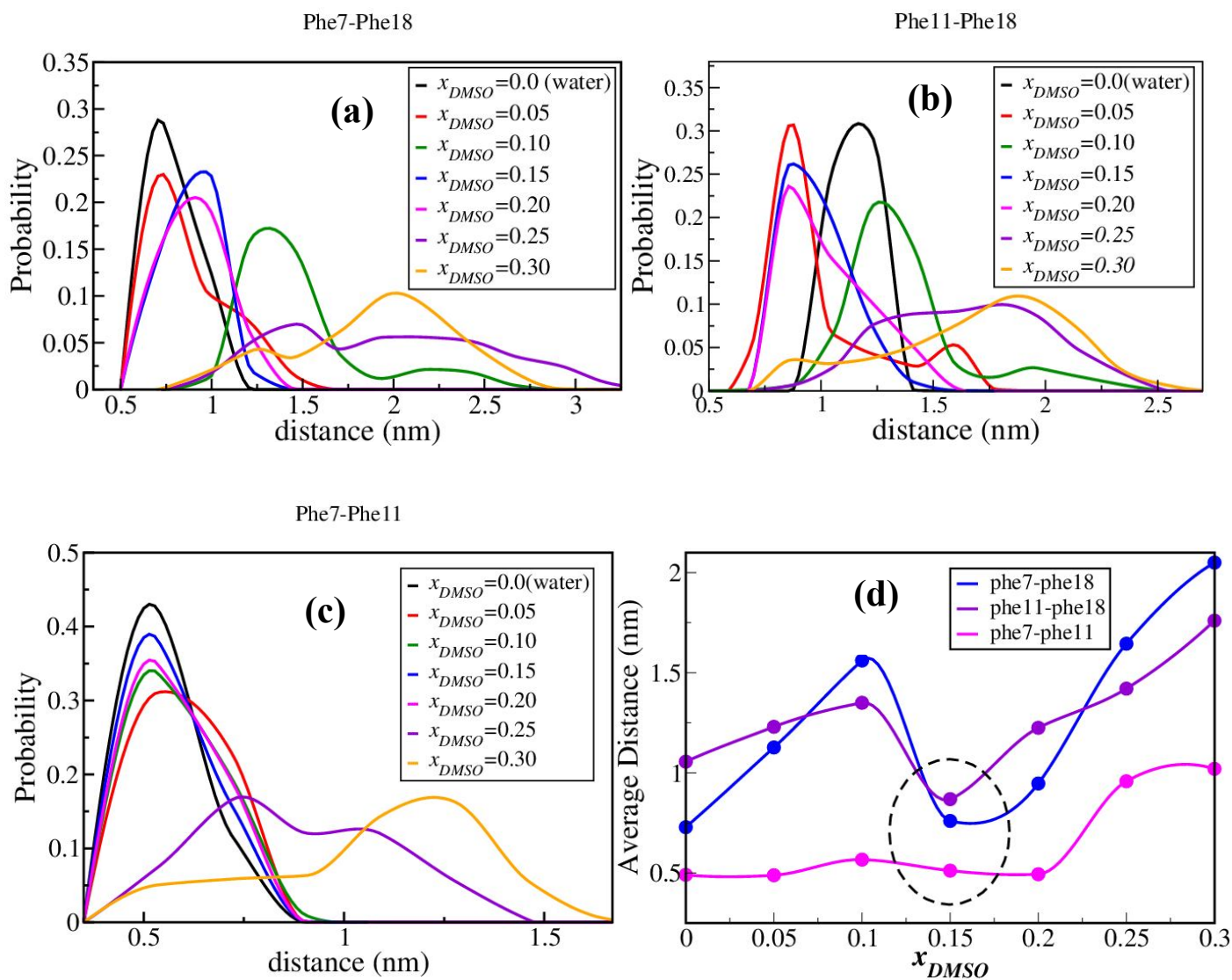

**Figure 11.** *Distribution of distances between residues of the hydrophobic core with increasing DMSO concentration. Probability distribution of distance between (a) phe-7 and phe-18 and (b) Phe-11 and Phe-18 (c) phe-7 and Phe-11(d) variation of average distance between these three pairs with changing DMSO concentration. The initial steps of separation of the hydrophobic core are found to be similar for both the water-ethanol and water-DMSO solutions at low co-solvent concentration. However with increase of ethanol concentration the protein is found to follow different pathways through formation of different intermediates in the two solutions.*

In order to compare the nature of structural changes of protein in water-DMSO with that in water-ethanol by varying co-solvent concentration, we plot distribution of distances of the groups forming the hydrophobic core in **Figure 11**(Phe-7-Phe18 and Phe-11-Phe18) at various DMSO concentrations. We find that on increasing DMSO concentration from $x_{DMSO} = 0.05$ to 0.10 the distance between both the pairs Phe7-Phe18 and Phe11-Phe18 increases, with the shift of distribution to a larger distance. On further increasing DMSO concentration ($x_{DMSO} = 0.15$) the distance between the hydrophobic pairs decrease to large extent, indicated by the movement of the distribution again towards a smaller distance. These results are in excellent agreement with that obtained for water-ethanol (**Figure 6**). We come to the conclusion that the initial steps of unfolding induced by co-solvents pursue the same path, which is the primary separation of the hydrophobic core followed by re-coalition of the same. However at further higher concentration the scenario takes different approach for the two cases. In water-DMSO, with progressive addition of DMSO, distance between the hydrophobic pairs start increasing leading to complete unfolding of the protein. We also plot comparative average distance between these three residues to perceive a prominent picture of the corresponding unfolding pathway.

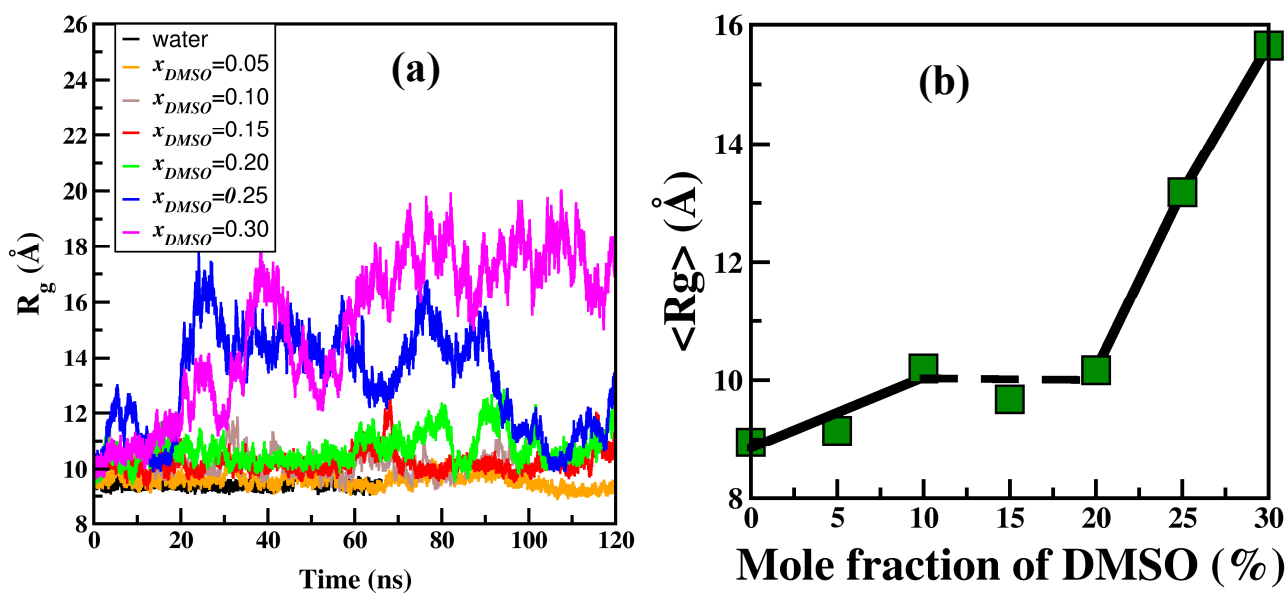

**Figure 12.** *Plots of variation of radius of gyration ($R_g$) for HP-36 in water-DMSO, taken from reference [37]. (a) Time evolution of radius of gyration of HP-36 in water-DMSO at various DMSO concentrations. (b) Average value of $R_g$ with increasing DMSO concentration. It can be seen that average $R_g$ value markedly increases with addition of DMSO, and finally leads to unfolding of the protein at $x_{DMSO} = 0.30$.*

On the other hand, on increasing ethanol concentration the protein again achieves a partially folded state ( $x_{eth} \approx 0.25$ ), with the signature of aggregation of hydrophobic core groups again, and never accomplish a fully unfolded state within the range of composition studied in this case. The same comparative signature is obtained from average radius of gyration (**Figure 12**), RMSD as well as equilibrium average of native contacts. Plot of these parameters for water-DMSO can be seen in reference [37].

# V. Theoretical analysis of solvent dependence of unfolding

Bryngelson-Wolynes theory of folding was developed in terms of two order parameters, namely, the fraction of native contacts, $\eta$, and the radius of gyration of the protein, $R_G$ [1, 2]. Corresponding free energy surface is characterized by multiple minima, with the deepest minima ideally corresponding to the native state under normal stability conditions of the folded protein. For example, for HP-36 at ambient conditions, the deepest minimum is the native state with η close to unity and the radius $R_G$ small, close to the compact folded configuration of the protein. As the protein is made to unfold, in the present cases through increase of DMSO or EtOH concentration in the binary mixtures, the free energy surface changes, with relative stability of the native state first decreasing with the increase of solute composition. The first major act of unfolding is the melting of the hydrophobic cluster formed by phenylalanines at 7, 11 and 18 positions. However, the details seem to depend critically on solvent conditions. In particular, the sequence of unfolding depends on relative depths of various minima and is hard to quantify. Solvent composition dependence allows one to probe the relative minima.

We have recently proposed a theoretical scheme to describe unfolding that combines aspects of Bryngelson-Wolynes theory with Marcus theory of electron transfer [37, 52-53]. The basic idea is to introduce (i) a set of local structural order parameters $\{\eta_i\}$ to describe formation of native secondary structures like helices and beta strands and (ii) another set of order parameters, $\{R_i\}$ to describe pair separation distances between amino acid residues that form tertiary native contacts in the folded state.

For HP-36, we may retain only three order parameters, $\eta_1$, $\eta_2$ and $\eta_3$ to describe formation of three helices. Simulations show that these three helices form and break at different stages.

Determination of the distance set $\{R_i\}$ is a bit more difficult. Simulations again show that at least for HP-36 only a few such distances are practically important, like the distance between Phe-11 and Phe-18, and that between Ala-9 and Leu35 [37]. One can thus describe the unfolding process in terms of two sets of order parameters: $\{\eta_i\}$ and $\{R_i\}$.

The advantage of introducing such an extended set of order parameters is that we can now describe relative stability of different secondary structures in terms of the relative minima of free energy, $F(\{\eta_i\}, \{R_i\})$. These minima are function of the solvent conditions. We can also now describe coupling between the two order parameters, within and across the set. Melting of HP-36 in water-DMSO mixture involves cooperative interaction between the breaking of helix 2 and separation between Phe7 and Phe18.

We next extend the two state model of protein folding/unfolding in terms of two state models for each of the order parameters, with minima located at $\{\eta_i^{UF}\}$ and $\{R_i^{UF}\}$ for the unfolded or extended state and $\{\eta_i^F\}$ and $\{R_i^F\}$ for the folded state, respectively.

Let us further define a probability distribution by $P(\{R_i\},\{\eta_i\},t)$, that has separation values, $\{R_i\}$ and the number of native contacts, $\{\eta_i\}$ at time $t$. The native state (F) is characterized by values $\{R_i^F\}$ and $\{\eta^F\}$.

We now follow the celebrated Marcus theory of electron transfer to incorporate essential features of the free energy surface in a rate constant [52, 53]. Let us consider the first step as separation between phe11 and phe18, accompanied by melting of the second helix. As these two order parameters ($\{R_i^F\}$, $\{\eta^F\}$) seem to be coupled and seem to vary in unison, we now consider only the separation $R_{11,18}$ as the order parameter to describe the initial stages of unfolding. The relevant free energy gap between the open state and associated compact state where phenylalanines are in contact is denoted by $\Delta G^0(\{R'\},\{\eta'\})$. The primes on $R$ and $\eta$ indicate that other values remain the same as in the native state. Marcus theory requires change in the value of the order parameter which is here equal to $R_{11,18} - a$, where $a$ is the separation in closed, native state. Marcus theory then provides the following expression for the activation energy of the initial melting process, [52]

$$E_{act} = \frac{\left[\Delta G^0(\{R'\},\{\eta'\}) + \lambda_{11,18}\right]^2}{4\lambda_{11,18}} \qquad (1)$$

where we assume,

$$\lambda_{11,18} = \frac{1}{2}\omega^2(R_{11,18} - a)^2 \qquad (2)$$

As mentioned, $\Delta G^0$ is the free energy gap between two minima when other values of the two order parameters are kept fixed at respective values in the native state and hence contains the effect of composition dependence. $\omega$ is the harmonic frequency of the free energy surface.

In neat water, the native state configuration is more stable. Thus, barrier is large. Also, the extended state is metatsable. So, even if it forms by fluctuation, it goes back to the native, compact state.

In case of water-DMSO binary mixture, as DMSO concentration increases, the unfolded state becomes more stable, and the barrier towards formation of this state decreases. At large $\Delta G^0$ (at $x_{DMSO} = 0$), we envisage a crossover to barrier less dynamics, as in Marcus theory of electron transfer.

We can now understand the difference between the DMSO and EtOH mixtures by using the above theory. Both the two species lead to stabilization of the unfolded state, albeit more for DMSO than for EtOH. But significant difference exists at larger composition. For DMSO, the other secondary structures also melt, as the extended or molten states gain more stability in this case. However, for EtOH the story is different. Here the molten or coiled state of the amino acid residues forming the 3$^{rd}$ helix, for example, never becomes more stable than the native state, in different EtOH concentrations (see schematic energy profile in **Figure 13**). Thus, it melts only partially.

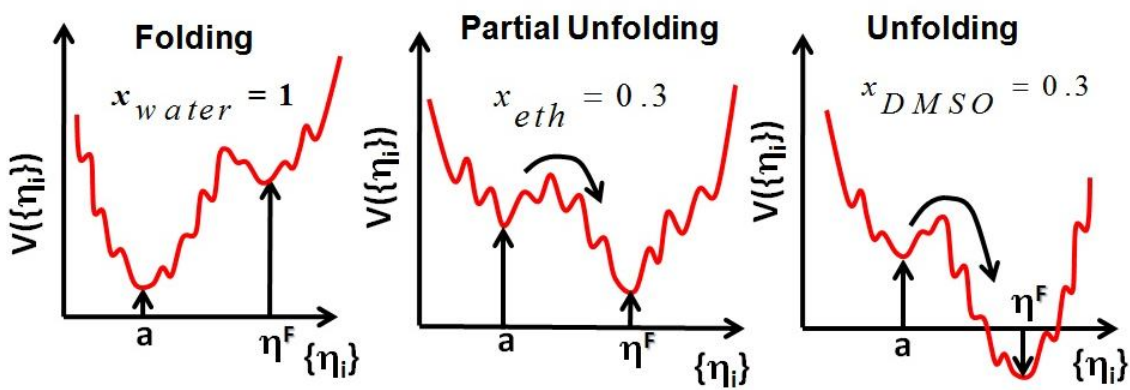

**Figure 13.** *Schematic energy profile of fraction of native contact involved in folding-unfolding transition. V({$\eta_i$}) represents the potential of mean force (PMF) where {$\eta_i$} are the distinct fractions of native contact order parameter . $\eta_i$*

**= a is the separation in native state while the $\eta_i = \eta_F$ gives the separation at the final state during unfolding. Note that in the case of ethanol the final state is a partially unfolded state. However, for DMSO the final stable state is a completely unfolded state. The above scheme is given for the two co-solvents when their concentration reaches to 0.30 mole fraction range.**

We now develop a simple theory to quantify the above discussions. As unfolding is a process much slower than solvent motion (or, any fast small amplitude motion of the protein), we can write down the following Smoluchowski equation for time evolution of the probability distribution, [54] $P\left(\{R_i\},\{\eta_i\},t\right)$:

$$\frac{\partial}{\partial t} P\left(\{R_i\},\{\eta_i\},t\right) = \sum_{i=1}^{N_{sp}} D_i \frac{1}{R_i^2} \frac{\partial}{\partial R_i} e^{-\beta V_i(\{R_i\},\{\eta_i\})} \frac{\partial}{\partial R_i} e^{\beta V_i(\{R_i\},\{\eta_i\})} P - \sum_{i=1}^{N_{ij}} \Gamma_i \eta_i \frac{\partial V(\{R_i\},\{\eta_i\})}{\partial \eta_i} \qquad (3)$$

Where we have assumed decoupling between the order parameters. This is certainly an approximation, but seems reasonable as a first step. However, the potential functions retain a dependence on other distances implicitly.

The potential functions, $V\left(\{R_i\},\{n_i\}\right)$ are to be regarded as potential of mean force (PMF) of the type described earlier [37]. Thus, in presence of DMSO, the energy surface, $V\left(R_{11,18},\eta_1,\eta_2\right)$, undergoes a sharp change compared to that of water or ethanol where a minimum at contact $\left(R_{11,18}\right)_{min} = a$ for mole fraction $x_{DMSO} = 0$ is replaced by a minimum at a larger distance. Therefore, unfolding at large denaturant concentration is in general a relaxation in a potential energy surface with a composition dependent activation barrier.

Adaptation of Marcus theory in the present case allows us to explain the onset of the nonequilibrium process of unfolding in terms of the increase in Gibbs' free energy gap, $\Delta G^0$ at different co-solvent environment. This in turn explores the tuning of the barrier height that holds the native state.

The progression of unfolding often begins with the step that induces, under appropriate conditions, subsequent separation of other contacts, signifying a high degree of cooperativity in the unfolding process [37]. For folding to unfolding transition of HP-36 the sequential steps of unfolding of helix-2 hold the distinct characteristics of global unfolding both in DMSO and ethanol. Nevertheless in both cases we compare the partial unfolding process of the full protein where at least the helix-2 undergoes complete unfolding. In order to observe the barrier sensitivity to the secondary structure of HP36 we evaluate the qualitative free energy landscape in ($\eta$, $R_g$) order plane. The figure shows the $2^{nd}$ helix melting process at $x_{DMSO}$=0.25 and $x_{EtOH}$=0.30. In case of DMSO the second helix melting process involves different stable intermediates ($I_D1$, $I_D2$) expressing two distinct minima separated by a barrier (>1 kJ/mol). On the contrary, in case of ethanol different intermediates ($I_E1$, $I_E2$ $I_E3$) evolve but separated by a shallow barrier (<0.4KJ/mol). This signifies that barrier separation of the intermediates direct the pathway of unfolding. Here for DMSO the barrier separation between the intermediates reveals that once it reaches to $I_D2$ from $I_D1$, it is unlikely that it can again go back towards $I_D1$. However for ethanol the small separation of barrier among the intermediates signifies that $I_E2$ has a significant probability to shift towards $I_E1$ or $I_E3$. Thus DMSO makes the second helix melting somewhat efficient than ethanol where the later makes several traps.

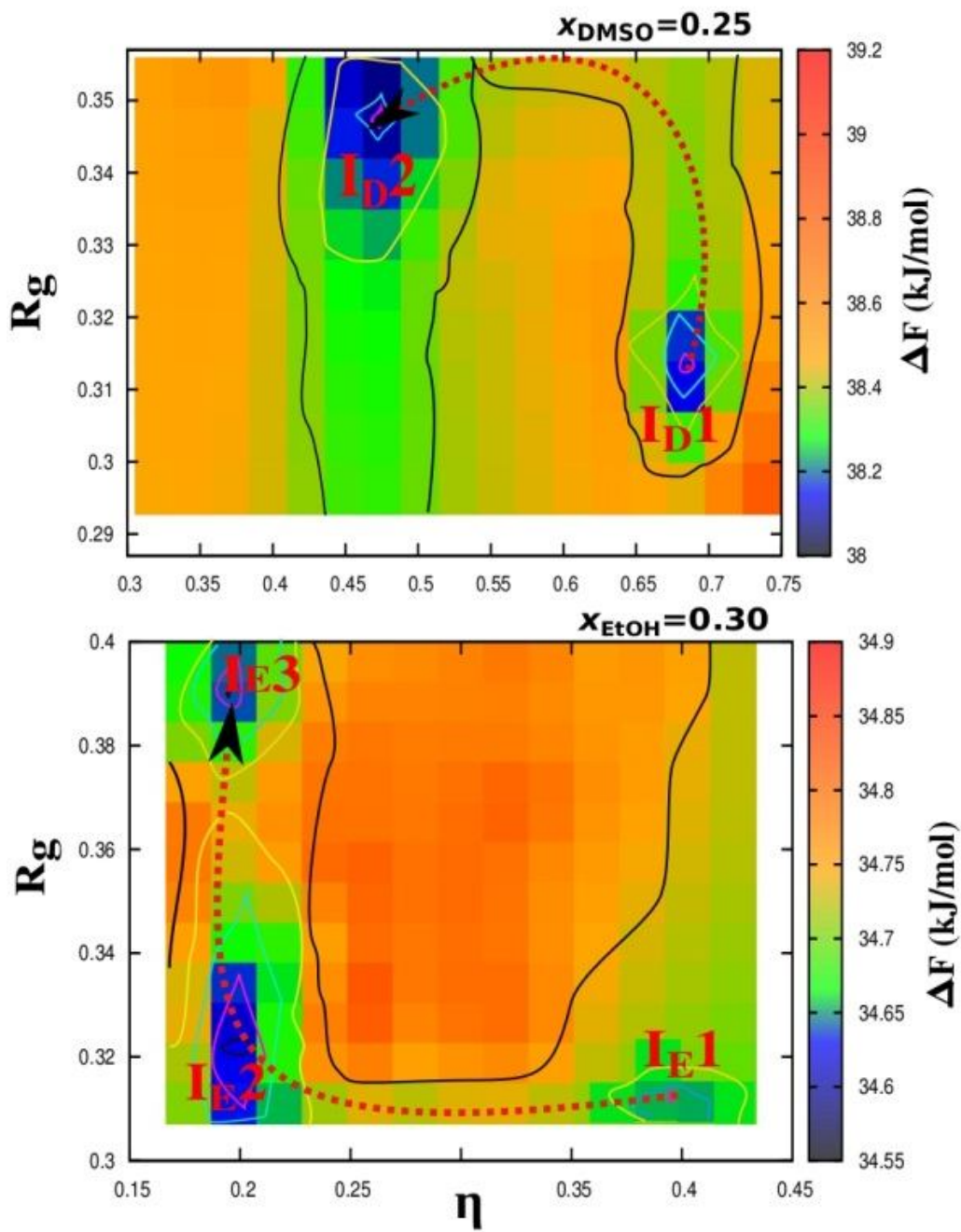

**Figure 14**. *Contour map of the two-order parameter (η, Rg) based free energy landscape of the secondary structure involving the 2nd helix of HP-36. For folding to unfolding transition of HP-36 the sequential step of unfolding of helix2 holds distinct characteristics of global unfolding both in DMSO and ethanol. In the two cases we compare the*

*partial unfolding process of the full protein where nothing but the helix-2 undergoes complete unfolding. Upper panel of the figure shows the 2ⁿᵈ helix melting process at $x_{DMSO}$=0.25. Presence of two distinct minima separated by a barrier corresponds to the emergence of different intermediates ($I_D1$, $I_D2$) from folding to unfolding pathway as indicted in the figure (arrowed line). The free energy separation (>1 kJ/mol) between two minima is evident from the energy landscape. The lower contour plot also demonstrates distinguishable unfolding path (arrowed line) evolving different intermediates ($I_E1$, $I_E2$ $I_E3$) but separated by a shallow barrier (<0.4KJ/mol). This plot shows the 2ⁿᵈ helix melting process at $x_{EtOH}$=0.30.*

# VI. Conclusion

The pathway of unfolding of a protein from stable native state to the extended unfolded (or, partially unfolded) state is often a complex one, comprising of unstable as well as partially stable intermediates which constitute the large number of local minima present in the free energy landscape. These intermediates are very hard to capture in general by means of experiments, because of their very short lifetimes. Recent developments in 2D-IR spectroscopy have allowed detecting beautifully the fast dynamics of proteins [3], and thus can be expected to identify the metastable intermediates formed in the unfolding pathway. The unfolding facilitated by different external factors may give rise to different intermediates. While thermal denaturation brings in deformation of the protein as well as modification of water structure that incorporate further conformational changes, chemical denaturation by using varied composition of co-solvents brings in static and dynamical changes in the protein by employing more subtle changes in intermolecular interactions among the protein and solvent(s).

In this work we study the effects of an important co-solvent ethanol on a 36 residue protein, chicken villin headpiece subdomain, commonly known as HP-36, and compare the results with that obtained for water-DMSO binary mixture. We vary the composition of the solvent and intend to see what type of dynamical changes in the structure of the protein take place. We indeed find that with change in ethanol concentration, considerable structural change is incorporated in the system. At low concentration of ethanol, partial unfolding of the protein occurs, accompanied by the signature of deformation of second helix. At $x_{eth} \approx 0.05 - 0.10$ we also find that the hydrophobic core groups (Phe-7, Phe11, Phe-18) are considerably separated from each other that primarily forms the basis of unfolding. In fact the initial pathway of unfolding is found to be similar with that of thermal denaturation as well as in water-DMSO. On increasing ethanol concentration further to $x_{eth} \approx 0.15 - 0.20$, we see that the structural change is not significant, and here we find that the largely separated hydrophobic contacts formed by the three phenylalanine groups start coming closer again. At $x_{eth} \approx 0.25$ surprisingly a native like state appears, with the consequent signatures in the $R_g$, RMSD values as well as in the average fraction of native contacts present in the structure. At this concentration, the hydrophobic core is associated in a similar way as that of native structure. This gives rise to a very unusual phenomenon, as we expect gradual unfolding of the protein with increasing ethanol concentration due to interaction between the hydrophobic groups of protein and co-solvent. After that on further increase of ethanol concentration different partially unfolded states of the protein are obtained and full denaturation is not accomplished even at ethanol concentration as high as $x_{eth} \approx 0.40$. Thus we show that although initiation of unfolding of HP-36 follows a universal path, further progress in the unfolding landscape may take on different routes depending on the environment. We compare the results with that obtained from unfolding study of the same protein in water-DMSO binary mixture, to show the similarity and differences in the pathway.

The reason for occurrence of such unusual phenomenon at around $x_{eth} \approx 0.25$ is attributed to the microheterogeneous phase separation in the water-ethanol solvent system itself that is brought about by the self-association of ethanol molecules through hydrogen bonding at that concentration. Evidence for such phenomena arising has been manifested by decreasing trend in diffusion co-efficient of ethanol as well as increasing number of ethanol-ethanol hydrogen bonds in that concentration range. We also show appearance of different partially unfolded states at different ethanol concentration, giving signature of multistage dynamics. These particular states signify various local minima in free energy landscape of protein-solvent system that are generated due to diverse mode of interactions between them. We have also provided a theoretical understanding of the phenomenon by employing Marcus theory combining with the vital aspects of Bryngelson-Wolynes theory. Our results essentially suggest that by tuning solvent concentration the dynamical evolution in the structure of protein can be altered significantly. This might effectively help in regulating enzymatic activity by controlling concentration of co-solvents, although further verification of the results is needed.

**ACKOWLEDGEMENT**

This article is dedicated to Professor Michel Fayer who has been a valued colleague and trusted friend to one of us (BB) for a many years. We wish Mike a long productive life with undiminished creative energy. This work was supported partly by grants from BRNS and DST. We thank JC Bose Fellowship for a partial support.

# References:

1. Bryngelson, D. J., Wolynes, G. P. *J. Phys. Chem.* **1989**, 93, 6902.

2. Bryngelson, J. D., Onuchic, J. N., Socci, N. D. & Wolynes, P. G. *Proteins Struct. Funct. Genet.* **1995**, 21, 167.

3. (a) Chung, J. K., Thielges, M. C., Fayer, M. D. *Proc. Natl. Acad. Sci. USA*. **2011**, 108, 3578. (b) Chung, J. K., Megan C. Thielges, M. C., Lynch, S. R., Fayer , M. D. *J. Phys. Chem. B*. **2012**, 116, 11024. (c) Thielges, M. C., Fayer, M. D. *Acc. Chem. Research*. **2012**, 45, 1866.

4. Kim, S., Chung, J. K., Kwak, K., Bren, K. L., Bagchi, B., Fayer, M. D. *J. Phys. Chem. B*. **2008**, 112, 10054.

5. Chung, H. S., Ganim, Z., Jones, K. C., Tokmakoff, A. *Proc. Natl. Acad. Sci. USA*. **2007**, 104, 14237.

6. Zanni, M.; Hochstrasser, R, M. *Curr. Opin. Struct. Biol.* **2001,** 11, 516.

7. Flory, P. J. *Principles of Polymer Chemistry (Cornell University Press: New York)* **1953**, 519.

8. Leopold, P. E., Montal, M. and Onuchic, J. N. *Proc. Natl. Acad. Sci. USA*. **1992,** 89, 8721.

9. Karplus, M. & Weaver, D.L. *Science*. **1976**, 260, 404.

10. Kim, P.S. & Baldwin, R.L. *Annu. Rev. Biochem*. **1982**, 51, 459.

11. Weissman, J.S. & Kim, P.S. *Science*. **1991**, 253, 1386.

12. Radford, S.E., Dobson, C.M. & Evans, P.A. *Nature*. **1992**, 358, 302.

13. Jackson, S.E. & Fersht, A.R. *Biochemistry*. **1991**, 30, 10428.

14. Koizumi, M.; Hirai, H.; Onai, T.; Inoue, K.; Hirai, M. *J. Appl. Cryst*. **2007**, 4, 175.

15. Li, W.; Zhou, R.; Mu, Y. *J. Phys. Chem. B* **2012**, 116, 1446.

16. Zhou, R. *PROTEINS: Structure, Function, and Genetics.* **2003**, 53, 148.

17. Levy Y., Onuchic JN. *Annu Rev Biophys Biomol Struct.* **2006**, 35, 389.

18. Levy, Y., Onuchic JN. *Proc. Natl. Acad. Sci. USA.* **2004**, 101, 3325.

19. Rhee, M. Y., Sorin, J. E., Jayachandran, G., Lindahl, E., Pande, S. V. *Proc. Natl. Acad. Sci. USA.* **2004**, 101, 6456.

20. Pizzitutti, F., Marchi, M., Sterpone, F., Rossky, P. J. *J. Phys. Chem. B.* **2007**, 111, 7584.

21. Jha, S.K., Dhar, D., Krishnamoorthy, G., Udgaonkar, J.B. *Proc. Natl. Acad. Sci. USA.* **2009**, 106, 11113.

22. Bhattacharya, K. *Chem. Commun.* **2008**, 2848.

23. Bagchi, B., Jana, B. *Chem. Soc. Rev.* **2010**, 39, 1936.

24. Chandra, A.; Bagchi, B.; *J. Phys. Chem.* **1991**, 95, 2529.

25.  Chandra, A.; Bagchi, B.; *J. Chem. Phys.* **1991**, 94, 8367.

26. (a) Roy, S.; Banerjee, S.; Biyani, N.; Jana, B.; Bagchi, B. *J. Phys. Chem. B.* **2010**, 115, 685, (b) Roy, S.; Banerjee, S. ; Bagchi, B. *J. Phys. Chem. B.* **2010**, 114, 12875.

27. Banerjee, S.; Ghosh, R.; Bagchi, B. *J. Phys. Chem. B.* **2012**, 116, 3713.

28. Juurinen, I.; Nakahara, K.; Ando, N.; Nishiumi, T.; Seta, H.; Yoshida, N.; Morinaga, T.; Itou, M.; Ninomiya, T.; Sakurai, Y.; Salonen, E.; Nordlund, K.; Hamalainen, K.; Hakala, M. *Phys. Rev. Lett.* **2011**, 107, 197401.

29. Yang, Z. W.; Tendian, S. W.; Carson, W. M.; Brouillette, W. J.; Delucas, L. *J. Protein Sci.*, **2004**, 13, 830.

30. Bhattacharjya, S.; Balaram, P. *Proteins.* **1997**, 29, 492.

31. Roy, S.; Jana, B.; Bagchi, B. *J. Chem. Phys.* 2012, 136, 115103.

32. Ortore, G. M.; Mariani, P.; Carsughi, F.; Cinelli, S.; Onori, G.; Teixeira, J.; Spinozzi, F. *J. Chem. Phys.* **2011**, 135, 245103.

33. Bhakkuni, V. *Archives of Biochemistry and Biophysics.,* **1998**, 357, 274.

34. Goodwin, A. C., Allen, T.J., Oslick, S. L., McClure K. F., Lee, J. H., Kemp, D. S., *J. Am. Chem. Soc.,* **1996**, 318, 3082.

35. Yoshida, K., *Nippon Kagakkai Koen Yokoshu.* 2005, 85, 376.

36. Yoshida, K.; Kawaguchi, J.; Lee, S.; Yamaguchi, T. *Pure Appl. Chem.,* **2008**, 80, 1337.

37. Roy, S.; Bagchi, B. *J. Phys. Chem. B.* **2013**, 117, 4488.

38. McKnight, C. J.; Matsudaira, P. T.; Kim, P. S. *Nat. Struct. Biol.* **1997**, 4, 180.

39. McKnight, C. J.; Doering, D. S.; Matsudaira, P. T.; Kim, P. S. *J. Mol. Biol.* **1996**, 260, 126.

40. Doering, D. S.; Matsudaira, P. *Biochemistry.* **1996**, 35, 12677.

41. Pope, B.; Way, M.; Matsudaira, P. T.; Weeds, A. *FEBS Lett.* **1994**, 338, 58.

42. Srinivas, G.; Bagchi, B. *J. Chem. Phys.* **2002**, 116, 8579.

43. Mukherjee, A.; Bagchi, B. *J. Chem. Phys.* **2003**, 118, 4733.

44. Bandyopadhyay, S.; Chakraborty, S.; Bagchi, B. *Phys. Chem. B.* **2004**, 108, 12608.

45. Berendsen, H. J. C.; Grigera, J. R.; Straatsma, T. P. *J. Phys. Chem.* **1987**, 91, 6269.

46. (a) Oostenbrink, C.; Villa, A.; Mark, A. E.; van Gunsteren, W. F. *J. Comput. Chem.* **2004**, 25, 1656. (b) Liu, H.; Mueller-Plathe, F.; van Gunsteren, W. F. *J. Am. Chem. Soc.* **1995**, 117,4363. (c) Geerke, D. P.; Oostenbrink, C.; van der Vegt, N. F. A.; van Gunsteren, W. F. *J. Phys. Chem. B.* **2004**, 108, 1436.

47. (a) Hoover, W. G. *Phys. Rev. A.* **1985**, 31, 1695. (b) Nose, S. *J. Chem. Phys.* **1984**, 81, 511.

48. Parrinello, M.; Rahman, A. *J. Appl. Phys.* **1981**, 52, 7182.

49. Frenkel, D.; Smit, B. *Understanding Molecular Simulation: From Algorithms to Applications*, 2nd ed.; Academic Press: San Diego, CA, **2002**.

50. Franks, F.; Ives, D. J. G. Q. *Rev. Chem. Soc.* **1966**, 20, 1.

51. Frank, H. S.; Evans, M. W. *J. Chem. Phys.* **1945**, 13, 507.

52. (a) Marcus, R. A. Annu. Rev. Phys. Chem. **1964**, 15, 155– 196. (b) Marcus, R. A. J. Chem. Phys. **1965**, 43, 679– 701. (c) Marcus, R. A.; Sutin, N. Biochim. Biophys. Acta **1985**, 811, 265– 322.

53. Sumi, H.; Marcus, R. A. J. Chem. Phys. **1986**, 84, 4894– 4914.

54. Bagchi, B. Molecular Relaxation in Liquids; Oxford University Press: New York, **2012**.